# β-Barium Borate (BBO) Absorption in the 0.188-6.22 μm Range

G. Tamošauskas

*Address:* Laser Research Center, Vilnius University, Saulėtekio ave. 10, LT10223 Vilnius, Lithuania

*E-mail:* gintaras.tamosauskas@ff.vu.lt

**Abstract:** The crystal of β-Barium Borate is widely used for nonlinear light wavelength conversion like harmonics generation and parametric amplification when pumped by the pulsed lasers and as electro-optically switched Pockels cells. Modern lasers produce hundreds of watts of average power, thus thermal effects become important. In this work, absorption of the crystal is obtained by reviewing published data, results of measurements performed by the author and data provided by the optics industry.

## 1. Introduction.

The β-Barium Borate (BBO) crystal, which was developed theoretically and synthesized by Chuangtian Chen and colleagues in 1970-80s [1, 2], is in constantly increasing demand for use with femtosecond lasers and has become standard in science and industry. Harmonics generators of such lasers and optical parametric amplifiers mostly rely on the use of BBO crystals. The crystal has a long and trusted use in nonlinear optics thanks to high nonlinearity and high optical damage threshold [3]. Its primary users were mostly owners of Ti:sapphire lasers which entered the market in 1990s and typically produce few watts of average power. The newer generation of femtosecond lasers based on directly-pumped Ytterbium-doped media produces hundreds of watts of average power. More importantly, such lasers are not unique scientific tools but mass-produced machinery ready for both scientific and industrial applications [4]. As a result, better characterization, simulation, and prediction of light wavelength manipulation-transformation processes should include thermal effects arising from absorption of fundamental laser wavelength and of harmonics where absorption coefficient could be significantly higher. See for example [5] how crystal detunes due to multi-photon absorption caused local heating. Optical parametric amplifiers lose their performance due to absorption in mid-infrared region. Pockels cells made of BBO crystals which are typically12-25 mm long can suffer due to heating especially when placed inside the laser cavity.

Over the years, attempts were made to measure and publish absorption coefficients of BBO crystals. At first, data was provided as transmittance graphs [6]. Later, high sensitivity thermal measurements were caried out [7]. The data is mostly available for lasers with approximately one micrometer fundamental wavelength and their harmonics. In the present work an attempt was made to review the published data, add new measurements, and generalize absorption coefficient covering the region from UV to mid-IR.

## 2. Methods and Results

The results are based on data available in publications and measurements carried out by the author. Two methods are typically used to acquire the absorption coefficient. The first one is based on transmittance measurement using a spectrophotometer. Later, data is recalculated taking into account reflections (using Fresnel law and Sellmeier equations), then absorption is calculated based on the length of the crystal. This method is accurate when volumetric absorption is high (typically above 3%). Otherwise, the result depends on the accuracy of Sellmeier equations, light polarization, and polarization ratio in the spectrophotometer as well as crystal orientation and its cut angle. Since Sellmeier equations were renewed recently [8] there are doubts regarding the accuracy of this method in earlier works for the UV and mid-IR regions where corrections are the largest. There is no data about test light polarization and crystal orientation in publications either. Therefore, the method is accurate just at the crystal spectral transmittance limits. It can be expected that even without knowing the polarization, an average value of Fresnel losses calculated for ordinary and extraordinary waves would provide a reliable result, but that is not true because birefringence of the BBO is very strong (difference in refractive indices is about 10%). Nevertheless, a lot of measurements have been carried out using different equipment and different length and cut angle crystals, whose results may represent typical tendencies [6, 8, 9-12].

The second method is based on heating the crystal with a powerful laser beam. In this case, temperature and thermal resistance (or amount of heat) should be known to determine the absorption coefficient which may be obtained directly or indirectly using probing beam [13]. This method allows to measure small values of absorption [7, 14-16] and is virtually not affected by the accuracy of the estimated reflections. The method is sensitive only to losses which result in crystal heating. The scattering and absorption which results in luminescence may not have an effect. Laser wavelength tuning is rather complicated as well when high power is necessary. Harmonics generators and optical parametric amplifiers can provide tunability of short pulses and this is exploited in numerous works.

The result of this work is presented in Fig. 1. The losses are presented as absorption or attenuation in power dB (absolute value) because it mostly originates from crystal heating. Intensity losses of the passing beam may be larger due to, for example, scattering [17]. The result is obtained by combining data in the literature, measurements made by author (UV and 1200-6220 nm IR) and UV to VIS data provided by EKSMA Optics Ltd. The circles in the figure are data points of the laser beam heating-based measurements (individual publications are not specified). The figure also includes CASTECH Inc. [18], the world's largest BBO producer's, minimal quality limits. Values are clipped at 150 dB/cm in UV and at 1000 dB/cm in mid-IR.

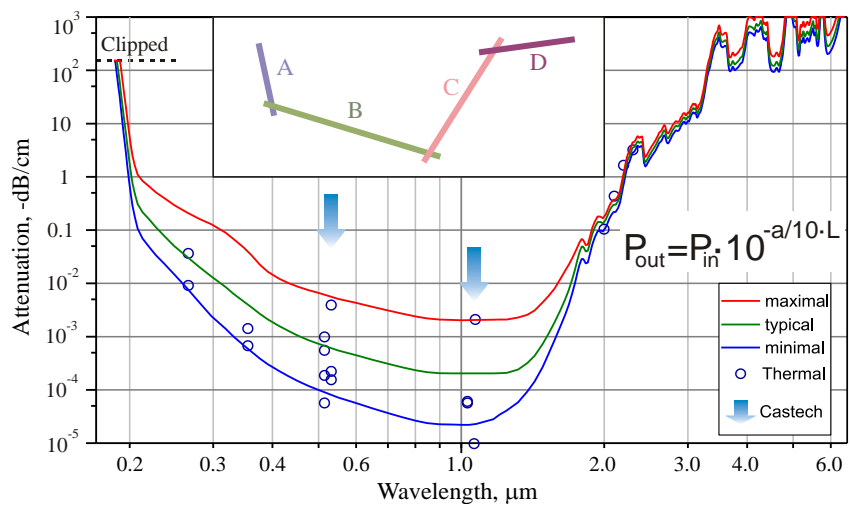

Fig. 1. Absorption losses of BBO.

The curve can be split into four regions (see insert). Region A defines UV absorption dictated by the crystal structure. Impurities also affect the region a lot. Absorption in Region C is defined by crystal structure similarly to Region A. However, influence of contaminants is much weaker, probably due to the absence of IR absorption of contaminants and grating defects which are by orders stronger in UV. Region D contains IR absorption of the crystal and overtones of the mid-far-IR absorption lines which are out of the present scale. Region B covers near UV, visible and near IR regions. It can be hypothesized that absorption here is defined by distant UV oscillator whose "tail" pervades Region B. Absorption level is very low and can be strongly affected by other factors. The region holds absorption of contaminants and crystal grating defects. Some samples showed increased absorption in 300 nm region, though its origin is not clear. General trend of absorption diminishing towards IR takes place, but values are scattered a lot from sample to sample as seen in published measurements. One can notice that the trend roughly follows $1/\lambda^4$ law. It is unlikely that scattering is responsible for the losses because measurements in the region are obtained using the heat monitoring method which is insensitive to light scattering.

The reliability of the measured absorption in different spectral regions should be discussed as well. There troubles with two existing definitions of absorption coefficient using natural logarithm or decimal logarithm (which historically came from intensity losses and power losses definitions). For example, optical glass manufacturers provide volumetric transmittance of 10 mm or 25 mm samples avoiding misinterpretation. Such a way to present the data does not seem attractive when loses are just few ppm/cm. In rear cases it is possible to understand which one is used if volumetric absorption is provided in parallel. Sometimes terms absorption and absorption coefficient are used as the same. It is assumed that majority of works, including thermal Common Pass Interferometry method, uses natural logarithm in definition of absorption coefficient. As for error bands, most reliable data in the IR region 1800-2700 nm and is consistent across various articles, resulting in a small uncertainty over the region. Data in the UV region (Region A) is much less uniform. There should be a fundamental limit for the minimal value defined purely by the crystal structure, but it is not known well. The expected maximal value includes overtop influence of contamination and defects, and can be rather high. Data in 250-1500 nm region is obtained by the thermal monitoring method only since there are no good alternatives to measure such small values. However, results in publications can differ by about two orders of magnitude. Therefore, the typical value may not be very clear. Data in the mid-IR region above 3.2 μm is obtained from a single publication where measurements of a thin crystal were made by two methods [8]. Data in the IR region 1200-6220 nm is based on spectrophotometer measurements for crystals cut at θ=20-35° angle.

Birefringent crystals may have different absorption pattern for ordinary and extraordinary light polarization. A single curve is provided in Fig. 1. Unfortunately, this aspect was left uninvestigated because it is mostly unknown how much light is polarized in the spectrophotometers (due to diffraction gratings). Most often it is unknown how crystals were oriented when measurements were caried out and often data on crystal cut angles was not available. Measurements made by the author of known cut angle crystals were performed in pairs by rotating crystals by 90 degrees. Later the processed data was averaged. Data obtained from the partners and from the literature often lack such information.

The result, including minimal and maximal bounds, is a free interpretation of the collected data. The explanation is as follows. The amount of data for the region around 250-1200 nm is small. There are no alternative simple and reliable techniques to define absorption of such a small value. Thermal measurements are quite complicated and of limited accuracy. Additionally, the method provides highly scattered values which may result from complicated crystal production. Absorption in the UV region highly depends on crystal or primary materials quality and provides values that differ a lot between samples. The result in mid-IR is obtained from a single source. Due to these obstacles, there is no possibility to provide statistically reliable data with proper account of each measurement's significance-weight factor. Therefore, the author made intuition-based interpolation of available data. Absorption losses presented as a typical value and as minimum and maximum values, which covers the majority (but not all) available experimental points.

The data is ready for use and provided in an easily accessible format (table) in Section 1 in Supplementary materials. To convert data into absorption coefficient for intensity losses $I_{out}=I_{in} \cdot \exp(-aL)$ a simple expression can be used $a = value(dB/cm)/10 \cdot \log_e(10)$. Unlike the Sellmeier equations, absorption data cannot be fitted by analytical function, which may cause inconvenience for certain calculations. Therefore, MATLAB® [19] and SCILAB [20] functions which return values of refractive indices and absorption losses and are capable of data interpolation are provided for public use in Section 2 and Section 3 in Supplementary materials. These functions can be used in dedicated environment or as an example for other programming-simulation environments.

**Conclusion**

BBO crystals are used across an extremely wide wavelength range. Some applications of nonlinear optics reach wavelengths close to UV and IR absorption edges where it is difficult to come up with a reasonable replacement. Other applications can provide hundreds of watts of average power for harmonics generation. The same is correct when BBO crystal is used as Pockels cell in the laser cavity. The data presented in this work allows to calculate typical absorption losses over wavelengths from UV-C to mid-IR, where the majority of applications take place. Absorption coefficient of the BBO crystal is provided as a table and can be easily used. Though absorption of a certain specimen can be rather different from the presented typical value, researchers and engineers can use the provided data to evaluate losses and stability of optical systems containing BBO crystals.

*Acknowledgements:*

Author thanks Mikhail Polyanskiy, the curator of the www.refractiveindex.info website, who kindly asked for the data of the BBO absorption together with renewed Sellmeier equations.

Special thanks to EKSMA Optics Ltd. for the data of transmittance measurements and in particular to Ramūnas Valauskas and Daugirdas Kuzma.

**Literature**

1. C. Chen, B. Wu, A. Jiang, G. You, "A new type ultraviolet SHG crystal - β-BaB$_2$O$_3$," Scientia Sinica, Ser. B **28**, p.235-243 (1985).
2. C. Chen, T. Sasaki, R. Li, Y. Wu, Z. Lin, Y. Mori, Z. Hu, J. Wang, G. Aka, M. Yoshimura, and Y. Kaneda, *Nonlinear optical borate crystals: principles and applications* (Wiley-WCH, 2012).

3. G.C. Bhar, A.K. Chaudhary, P. Kumbhakar, A.M. Rudra, S.C. Sabarwal, "A comparative study of laser-induced surface damage thresholds in BBO crystals and effect of impurities," Opt. Materials **27**, 119–123 (2004).

4. https://lightcon.com/product/carbide-femtosecond-lasers/ (viewed on 2021 10 06)

5. E. Gabryte, S. Sobutas, M. Vengris, R. Danielius, "Control of thermal effects in fast-switching femtosecond UV laser," Appl. Phys. B **120**, 31–39 (2015), DOI 10.1007/s00340-015-6090-4

6. D. Eimerl, L. Davis, S. Velsko, E. K. Graham, and A. Zalkin, "Optical, mechanical, and thermal properties of barium borate," J. Appl. Phys. **62**, p. 1968–1983 (1987).

7. D. Perlov, S. Livneh, P. Czechowicz, A. Goldgirsh, and D. Loiacono, "Progress in growth of large β-$BaB_2O_4$ single crystals," Cryst. Res. Technol. **46**, No. 7, p. 651 – 654 (2011), DOI: 10.1002/crat.201100208

8. G. Tamošauskas, G. Beresnevičius, D. Gadonas, and A. Dubietis, "Transmittance and phase matching of BBO crystal in the 3−5 μm range and its application for the characterization of mid-infrared laser pulses," Opt. Mater. Express **8**, p. 1410-1418 (2018).

9. C. L. Tang, "*Final Report to NRL on* Growth, Characterization, and Applications of β-Barium Borate and Related Crystals," (1994)

10. G.C.Bhar, U.Chatterjee, A.M.Rudra, P.Kumbhakar, "Tunable coherent far-UVgeneration by frequency conversion in BBO," Quantum Electron. **29** 800, p. 245-250 (1999).

11. R. Bhandari, T. Taira, A. Miyamoto, Y. Furukawa, and T. Tago, "> 3 MW peak power at 266 nm using Nd:YAG/$Cr^{4+}$:YAG microchip laser and fluxless-BBO," Opt. Material Express **2**, No. 7, p. 907-913 (2012).

12. L. Deyra, A. Maillard, R. Maillard, D. Sangla, F. Salin, F. Balembois, A.E. Kokh, P. Georges, "Impact of $BaB_2O_4$ growth method on frequency conversion to the deep ultra-violet," Solid State Sciences (2015), DOI: 10.1016/j.solidstatesciences.2015.10.019

13. A. Alexandrovski, Martin Fejer, A. Markosian, Roger Route, "Photothermal common-path interferometry (PCI): new developments," Proc. SPIE 7193, Solid State Lasers XVIII: Technology and Devices, 71930D (28 February 2009); DOI: 10.1117/12.814813

14. R. Riedel, J. Rothhardt, K. Beil, B. Gronloh, A. Klenke, H. Höppner, M. Schulz, U. Teubner, C. Kränkel, J. Limpert, A. Tünnermann, M.J. Prandolini, and F. Tavella, "Thermal properties of borate crystals for high power optical parametric chirped-pulse amplification," Optics Express **22**, p. 17607-17619 (2014), DOI:10.1364/OE.22.017607

15. M. Stubenvoll, B. Schäfer, K. Mann, and O. Novak, "Photothermal method for absorption measurements in anisotropic crystals," Review of Scientific Instruments **87**, 023904 (2016), DOI: 10.1063/1.4941662

16. Ch. Mühlig, S. Bublitz, "Characterization of NLO crystal absorption for wavelengths 1ω to 4ω," Proc. SPIE 10014, Laser-Induced, Damage in Optical Materials 2016, 100141N (6 December 2016), DOI: 10.1117/12.2245023

17. H. Kouta and Y. Kuwano, "Annealing to reduce scattering centers in Czochralski-grown β-$BaB_2O_4$," Appl. Opt. **38**, p. 1053-1057 (1999).

18. https://www.castech.com/product/BBO-120.html (viewed on 2021 10 10)

19. www.mathworks.com (viewed on 2021 10 06)

20. www.scilab.org (viewed on 2021 10 06)

# Supplementary materials.

**Section 1.**

Table (list) of BBO absorption losses in power -**dB/cm** (absolute value is provided).

Wavelength is provided in µm. Absorption losses for length of 10 mm (cm$^{-1}$). $P_{out}=P_{in} \cdot 10^{-a/10 \cdot L}$
Example: a=20 dB/cm, L=0.5 cm, Pin=1 W; $P_{out}=P_{in} \cdot 10^{-a/10 \cdot L}= 1*10^{\wedge}(-20/10*0.5)=10^{\wedge}(-1)=0.1W$.
The table is designed for copy and paste form the PDF document. Space delimited.
Tips for Excel users: after copying data press *Paste* icon roll down > *Use text import wizard* > *Finish*

| Wavelength | Min | Typical | Max | Wavelength | Min | Typical | Max | Wavelength | Min | Typical | Max |
|---|---|---|---|---|---|---|---|---|---|---|---|
| 1.8500E-01 | 1.5000E+02 | 1.5000E+02 | 1.5000E+02 | 2.8728E+00 | 7.6809E+00 | 9.3095E+00 | 1.1283E+01 | 4.4088E+00 | 3.3447E+02 | 4.3481E+02 | 6.5221E+02 |
| 1.8530E-01 | 1.5000E+02 | 1.5000E+02 | 1.5000E+02 | 2.8824E+00 | 8.2230E+00 | 9.9542E+00 | 1.2050E+01 | 4.4171E+00 | 2.5545E+02 | 3.3209E+02 | 4.9813E+02 |
| 1.8590E-01 | 1.5000E+02 | 1.5000E+02 | 1.5000E+02 | 2.8901E+00 | 8.7495E+00 | 1.0580E+01 | 1.2794E+01 | 4.4238E+00 | 2.1344E+02 | 2.7747E+02 | 4.1621E+02 |
| 1.8640E-01 | 1.3157E+02 | 1.5000E+02 | 1.5000E+02 | 2.8992E+00 | 9.1253E+00 | 1.1022E+01 | 1.3313E+01 | 4.4305E+00 | 1.7919E+02 | 2.3295E+02 | 3.4942E+02 |
| 1.8679E-01 | 1.1200E+02 | 1.5000E+02 | 1.5000E+02 | 2.9101E+00 | 9.6453E+00 | 1.1637E+01 | 1.4039E+01 | 4.4421E+00 | 1.4275E+02 | 1.8557E+02 | 2.7836E+02 |
| 1.8701E-01 | 1.0324E+02 | 1.5000E+02 | 1.5000E+02 | 2.9212E+00 | 1.0331E+01 | 1.2450E+01 | 1.5004E+01 | 4.4604E+00 | 1.1618E+02 | 1.5104E+02 | 2.2656E+02 |
| 1.8745E-01 | 8.6507E+01 | 1.5000E+02 | 1.5000E+02 | 2.9329E+00 | 1.0774E+01 | 1.2970E+01 | 1.5615E+01 | 4.4796E+00 | 1.0112E+02 | 1.3145E+02 | 1.9718E+02 |
| 1.8810E-01 | 6.7513E+01 | 1.2760E+02 | 1.5000E+02 | 2.9421E+00 | 1.0783E+01 | 1.2970E+01 | 1.5602E+01 | 4.4971E+00 | 9.6166E+01 | 1.2502E+02 | 1.8752E+02 |
| 1.8870E-01 | 5.4510E+01 | 1.0329E+02 | 1.5000E+02 | 2.9520E+00 | 1.0507E+01 | 1.2630E+01 | 1.5182E+01 | 4.5179E+00 | 9.3516E+01 | 1.2157E+02 | 1.8236E+02 |
| 1.8930E-01 | 4.3283E+01 | 8.2200E+01 | 1.5000E+02 | 2.9602E+00 | 9.8323E+00 | 1.1813E+01 | 1.4191E+01 | 4.5297E+00 | 9.2997E+01 | 1.2090E+02 | 1.8135E+02 |
| 1.9021E-01 | 3.0908E+01 | 5.8907E+01 | 1.5000E+02 | 2.9725E+00 | 9.9276E+00 | 1.1921E+01 | 1.4315E+01 | 4.5381E+00 | 9.2997E+01 | 1.2090E+02 | 1.8135E+02 |
| 1.9100E-01 | 2.2154E+01 | 4.2600E+01 | 1.2808E+02 | 2.9795E+00 | 1.0446E+01 | 1.2540E+01 | 1.5053E+01 | 4.5520E+00 | 9.3516E+01 | 1.2157E+02 | 1.8236E+02 |
| 1.9150E-01 | 1.7920E+01 | 3.5000E+01 | 1.0553E+02 | 2.9907E+00 | 1.1112E+01 | 1.3336E+01 | 1.6005E+01 | 4.5651E+00 | 9.7166E+01 | 1.2632E+02 | 1.8947E+02 |
| 1.9204E-01 | 1.3984E+01 | 2.8198E+01 | 8.5532E+01 | 2.9998E+00 | 1.1113E+01 | 1.3336E+01 | 1.6004E+01 | 4.5774E+00 | 1.0221E+02 | 1.3288E+02 | 1.9932E+02 |
| 1.9399E-01 | 6.2526E+00 | 1.3500E+01 | 4.1499E+01 | 3.0093E+00 | 1.0822E+01 | 1.2987E+01 | 1.5586E+01 | 4.5842E+00 | 1.0564E+02 | 1.3733E+02 | 2.0599E+02 |
| 1.9597E-01 | 2.8122E+00 | 6.5564E+00 | 2.0477E+01 | 3.0184E+00 | 1.1730E+01 | 1.4077E+01 | 1.6898E+01 | 4.5938E+00 | 1.0930E+02 | 1.4209E+02 | 2.1314E+02 |
| 1.9746E-01 | 1.5182E+00 | 3.7480E+00 | 1.1854E+01 | 3.0273E+00 | 1.2639E+01 | 1.5170E+01 | 1.8215E+01 | 4.6058E+00 | 1.1052E+02 | 1.4367E+02 | 2.1551E+02 |
| 1.9913E-01 | 8.0279E-01 | 2.0557E+00 | 6.5599E+00 | 3.0369E+00 | 1.3715E+01 | 1.6464E+01 | 1.9778E+01 | 4.6156E+00 | 1.0927E+02 | 1.4204E+02 | 2.1307E+02 |
| 2.0059E-01 | 4.7717E-01 | 1.2426E+00 | 3.9932E+00 | 3.0484E+00 | 1.4776E+01 | 1.7742E+01 | 2.1328E+01 | 4.6227E+00 | 1.0510E+02 | 1.3663E+02 | 2.0495E+02 |
| 2.0241E-01 | 2.8270E-01 | 7.4901E-01 | 2.4441E+00 | 3.0598E+00 | 1.5522E+01 | 1.8643E+01 | 2.2430E+01 | 4.6295E+00 | 1.0161E+02 | 1.3210E+02 | 1.9814E+02 |
| 2.0459E-01 | 1.7431E-01 | 4.7702E-01 | 1.6080E+00 | 3.0709E+00 | 1.6242E+01 | 1.9514E+01 | 2.3500E+01 | 4.6367E+00 | 9.8238E+01 | 1.2771E+02 | 1.9156E+02 |
| 2.0748E-01 | 1.0629E-01 | 3.0757E-01 | 1.0975E+00 | 3.0801E+00 | 1.6847E+01 | 1.8847E+01 | 2.2721E+01 | 4.6481E+00 | 9.3920E+01 | 1.2210E+02 | 1.8314E+02 |
| 2.1239E-01 | 6.9493E-02 | 2.1715E-01 | 8.4060E-01 | 3.0895E+00 | 1.5373E+01 | 1.8483E+01 | 2.2307E+01 | 4.6589E+00 | 9.2884E+01 | 1.2075E+02 | 1.8112E+02 |
| 2.1978E-01 | 4.6592E-02 | 1.5449E-01 | 6.4696E-01 | 3.1000E+00 | 1.6009E+01 | 1.9256E+01 | 2.3267E+01 | 4.6688E+00 | 9.7666E+01 | 1.2697E+02 | 1.9045E+02 |
| 2.2664E-01 | 3.3634E-02 | 1.1258E-01 | 5.0944E-01 | 3.1080E+00 | 1.5942E+01 | 1.9182E+01 | 2.3204E+01 | 4.6754E+00 | 1.0273E+02 | 1.3355E+02 | 2.0033E+02 |
| 2.3555E-01 | 2.3632E-02 | 7.8577E-02 | 4.0108E-01 | 3.1139E+00 | 1.4885E+01 | 1.7916E+01 | 2.1697E+01 | 4.6844E+00 | 1.1008E+02 | 1.4310E+02 | 2.1466E+02 |
| 2.4777E-01 | 1.4629E-02 | 4.9864E-02 | 2.9194E-01 | 3.1237E+00 | 1.3412E+01 | 1.6151E+01 | 1.9587E+01 | 4.6924E+00 | 1.1867E+02 | 1.5427E+02 | 2.3140E+02 |
| 2.6379E-01 | 7.9651E-03 | 3.1086E-02 | 2.1196E-01 | 3.1327E+00 | 1.2759E+01 | 1.5373E+01 | 1.8673E+01 | 4.7014E+00 | 1.2704E+02 | 1.6515E+02 | 2.4773E+02 |
| 2.8009E-01 | 4.3116E-03 | 1.9992E-02 | 1.6043E-01 | 3.1445E+00 | 1.3656E+01 | 1.6464E+01 | 2.0037E+01 | 4.7074E+00 | 1.3609E+02 | 1.7691E+02 | 2.6537E+02 |
| 2.9662E-01 | 2.4107E-03 | 1.3427E-02 | 1.2846E-01 | 3.1550E+00 | 1.4884E+01 | 1.7958E+01 | 2.1900E+01 | 4.7178E+00 | 1.4990E+02 | 1.9487E+02 | 2.9230E+02 |
| 3.1798E-01 | 1.3960E-03 | 8.4544E-03 | 8.8707E-02 | 3.1670E+00 | 1.5624E+01 | 1.8865E+01 | 2.3060E+01 | 4.7270E+00 | 1.6373E+02 | 2.1285E+02 | 3.1927E+02 |
| 3.3000E-01 | 1.0057E-03 | 6.3708E-03 | 6.8134E-02 | 3.1790E+00 | 1.6482E+01 | 1.9917E+01 | 2.4403E+01 | 4.7397E+00 | 1.8537E+02 | 2.4097E+02 | 3.6146E+02 |
| 3.4702E-01 | 6.8104E-04 | 4.5391E-03 | 4.6718E-02 | 3.1878E+00 | 1.7514E+01 | 2.1179E+01 | 2.6003E+01 | 4.7486E+00 | 2.1561E+02 | 2.8029E+02 | 4.2044E+02 |
| 3.6000E-01 | 5.2113E-04 | 3.5478E-03 | 3.3776E-02 | 3.1964E+00 | 1.9249E+01 | 2.3294E+01 | 2.8660E+01 | 4.7576E+00 | 2.5227E+02 | 3.2795E+02 | 4.9193E+02 |
| 3.7556E-01 | 3.8512E-04 | 2.6533E-03 | 2.2261E-02 | 3.2068E+00 | 2.1564E+01 | 2.6117E+01 | 3.2211E+01 | 4.7684E+00 | 2.9868E+02 | 3.8828E+02 | 5.8242E+02 |
| 3.9000E-01 | 3.0126E-04 | 2.0939E-03 | 1.6431E-02 | 3.2176E+00 | 2.4681E+01 | 2.9919E+01 | 3.7002E+01 | 4.7747E+00 | 3.4407E+02 | 4.4729E+02 | 6.7093E+02 |
| 4.1156E-01 | 2.2589E-04 | 1.5837E-03 | 1.2194E-02 | 3.2316E+00 | 2.9401E+01 | 3.5680E+01 | 4.4268E+01 | 4.7812E+00 | 3.8903E+02 | 5.0573E+02 | 7.5860E+02 |
| 4.3000E-01 | 1.8176E-04 | 1.2859E-03 | 1.0158E-02 | 3.2435E+00 | 3.5235E+01 | 4.2805E+01 | 5.3272E+01 | 4.7875E+00 | 4.5130E+02 | 5.8668E+02 | 8.8003E+02 |
| 4.5088E-01 | 1.4566E-04 | 1.0425E-03 | 8.6347E-03 | 3.2540E+00 | 4.1919E+01 | 5.0973E+01 | 6.3613E+01 | 4.7952E+00 | 5.3904E+02 | 7.0855E+02 | 1.0000E+03 |
| 4.7500E-01 | 1.1791E-04 | 8.5507E-04 | 7.4400E-03 | 3.2615E+00 | 4.8339E+01 | 5.8827E+01 | 7.7390E+01 | 4.8010E+00 | 6.0767E+02 | 7.8997E+02 | 1.0000E+03 |
| 4.9996E-01 | 9.8396E-05 | 7.2497E-04 | 6.5372E-03 | 3.2704E+00 | 5.6486E+01 | 6.8802E+01 | 8.6282E+01 | 4.8066E+00 | 6.8462E+02 | 8.9000E+02 | 1.0000E+03 |
| 5.4934E-01 | 7.1447E-05 | 5.4365E-04 | 5.0791E-03 | 3.2801E+00 | 6.4664E+01 | 7.8836E+01 | 9.9135E+01 | 4.8092E+00 | 7.2329E+02 | 9.4028E+02 | 1.0000E+03 |
| 5.9999E-01 | 5.5989E-05 | 4.4213E-04 | 4.2683E-03 | 3.2896E+00 | 7.6027E+01 | 9.2780E+01 | 1.1700E+02 | 4.8120E+00 | 7.5708E+02 | 9.8421E+02 | 1.0000E+03 |
| 6.4824E-01 | 4.4599E-05 | 3.6501E-04 | 3.6058E-03 | 3.3007E+00 | 8.7526E+01 | 1.0693E+02 | 1.3525E+02 | 4.8161E+00 | 8.1538E+02 | 1.0000E+03 | 1.0000E+03 |
| 7.0098E-01 | 3.6078E-05 | 3.0532E-04 | 3.0707E-03 | 3.3098E+00 | 9.5709E+01 | 1.1704E+02 | 1.4847E+02 | 4.8226E+00 | 9.2308E+02 | 1.0000E+03 | 1.0000E+03 |
| 7.5036E-01 | 3.0249E-05 | 2.6250E-04 | 2.6652E-03 | 3.3196E+00 | 1.0303E+02 | 1.2612E+02 | 1.6045E+02 | 4.8300E+00 | 1.0000E+03 | 1.0000E+03 | 1.0000E+03 |
| 7.9973E-01 | 2.6049E-05 | 2.3197E-04 | 2.3728E-03 | 3.3279E+00 | 1.1013E+02 | 1.3493E+02 | 1.7215E+02 | 4.8400E+00 | 1.0000E+03 | 1.0000E+03 | 1.0000E+03 |
| 8.4462E-01 | 2.3588E-05 | 2.1362E-04 | 2.1897E-03 | 3.3377E+00 | 1.2529E+02 | 1.5367E+02 | 1.9663E+02 | 4.8587E+00 | 1.0000E+03 | 1.0000E+03 | 1.0000E+03 |
| 8.9624E-01 | 2.2470E-05 | 2.0499E-04 | 2.0851E-03 | 3.3467E+00 | 1.4354E+02 | 1.7625E+02 | 2.2624E+02 | 4.9097E+00 | 1.0000E+03 | 1.0000E+03 | 1.0000E+03 |
| 9.5571E-01 | 2.2184E-05 | 2.0358E-04 | 2.0407E-03 | 3.3595E+00 | 1.6512E+02 | 2.0302E+02 | 2.6158E+02 | 4.9400E+00 | 1.0000E+03 | 1.0000E+03 | 1.0000E+03 |
| 1.0017E+00 | 2.1964E-05 | 2.0280E-04 | 2.0177E-03 | 3.3720E+00 | 1.8921E+02 | 2.3295E+02 | 3.0130E+02 | 4.9586E+00 | 1.0000E+03 | 1.0000E+03 | 1.0000E+03 |
| 1.0350E+00 | 2.1884E-05 | 2.0280E-04 | 2.0235E-03 | 3.3821E+00 | 2.1581E+02 | 2.6602E+02 | 3.4532E+02 | 4.9756E+00 | 8.4615E+02 | 1.0000E+03 | 1.0000E+03 |
| 1.0714E+00 | 2.2734E-05 | 2.0290E-04 | 2.0415E-03 | 3.3936E+00 | 2.3823E+02 | 2.9405E+02 | 3.8313E+02 | 4.9832E+00 | 7.7692E+02 | 1.0000E+03 | 1.0000E+03 |
| 1.1651E+00 | 2.7907E-05 | 2.0384E-04 | 2.0681E-03 | 3.4057E+00 | 2.6325E+02 | 3.2537E+02 | 4.2555E+02 | 4.9875E+00 | 7.4902E+02 | 9.7372E+02 | 1.0000E+03 |
| 1.2000E+00 | 3.3916E-05 | 2.0511E-04 | 2.0859E-03 | 3.4166E+00 | 2.8597E+02 | 3.5389E+02 | 4.6450E+02 | 4.9935E+00 | 7.1784E+02 | 9.3319E+02 | 1.0000E+03 |
| 1.2426E+00 | 4.0699E-05 | 2.0689E-04 | 2.0794E-03 | 3.4253E+00 | 2.9751E+02 | 3.6857E+02 | 4.8527E+02 | 5.0023E+00 | 6.7369E+02 | 8.7583E+02 | 1.0000E+03 |
| 1.3130E+00 | 5.8122E-05 | 2.3855E-04 | 2.2704E-03 | 3.4346E+00 | 3.0556E+02 | 3.7894E+02 | 5.0036E+02 | 5.0212E+00 | 6.3493E+02 | 8.2553E+02 | 1.0000E+03 |
| 1.3735E+00 | 9.2663E-05 | 3.1658E-04 | 2.5534E-03 | 3.4410E+00 | 3.1018E+02 | 3.8501E+02 | 5.0964E+02 | 5.0391E+00 | 6.4659E+02 | 8.4088E+02 | 1.0000E+03 |
| 1.4233E+00 | 1.4307E-04 | 4.3615E-04 | 3.0754E-03 | 3.4488E+00 | 3.2169E+02 | 3.9967E+02 | 5.3042E+02 | 5.0545E+00 | 6.5791E+02 | 8.5583E+02 | 1.0000E+03 |
| 1.4786E+00 | 2.6365E-04 | 7.1681E-04 | 4.3422E-03 | 3.4580E+00 | 3.3522E+02 | 4.1687E+02 | 5.5472E+02 | 5.0638E+00 | 6.1900E+02 | 8.0540E+02 | 1.0000E+03 |
| 1.5218E+00 | 4.6930E-04 | 1.1256E-03 | 5.7325E-03 | 3.4630E+00 | 3.5696E+02 | 4.4424E+02 | 5.9241E+02 | 5.0680E+00 | 5.5329E+02 | 7.2002E+02 | 1.0000E+03 |
| 1.5681E+00 | 8.6597E-04 | 1.8544E-03 | 7.9829E-03 | 3.4697E+00 | 3.8588E+02 | 4.8062E+02 | 6.4236E+02 | 5.0739E+00 | 4.7401E+02 | 6.1696E+02 | 9.2640E+02 |
| 1.6139E+00 | 1.5595E-03 | 3.1450E-03 | 1.2130E-02 | 3.4775E+00 | 4.0438E+02 | 5.0410E+02 | 6.7539E+02 | 5.0800E+00 | 3.7938E+02 | 4.9389E+02 | 7.4175E+02 |
| 1.6517E+00 | 2.5993E-03 | 5.0696E-03 | 1.7459E-02 | 3.4840E+00 | 4.1511E+02 | 5.1791E+02 | 6.9555E+02 | 5.0866E+00 | 3.1395E+02 | 4.0881E+02 | 6.1411E+02 |
| 1.6798E+00 | 3.9662E-03 | 7.5127E-03 | 2.2440E-02 | 3.4919E+00 | 3.9540E+02 | 4.9379E+02 | 6.6489E+02 | 5.0952E+00 | 2.5389E+02 | 3.3072E+02 | 4.9695E+02 |
| 1.7056E+00 | 6.2740E-03 | 1.1536E-02 | 2.8993E-02 | 3.5010E+00 | 3.4233E+02 | 4.2795E+02 | 5.7789E+02 | 5.1084E+00 | 2.0633E+02 | 2.6891E+02 | 4.0425E+02 |
| 1.7318E+00 | 1.0280E-02 | 1.8379E-02 | 3.8158E-02 | 3.5089E+00 | 3.0595E+02 | 3.8287E+02 | 5.1853E+02 | 5.1210E+00 | 1.8641E+02 | 2.4309E+02 | 3.6563E+02 |
| 1.7493E+00 | 1.5030E-02 | 2.6322E-02 | 4.6357E-02 | 3.5182E+00 | 2.8087E+02 | 3.5193E+02 | 4.7826E+02 | 5.1351E+00 | 1.8826E+02 | 2.4568E+02 | 3.6974E+02 |
| 1.7687E+00 | 2.2052E-02 | 3.8021E-02 | 5.8097E-02 | 3.5341E+00 | 2.7013E+02 | 3.3902E+02 | 4.6280E+02 | 5.1482E+00 | 2.2679E+02 | 2.9615E+02 | 4.4595E+02 |
| 1.7878E+00 | 2.8513E-02 | 4.8691E-02 | 6.7548E-02 | 3.5461E+00 | 2.6841E+02 | 3.3741E+02 | 4.6266E+02 | 5.1550E+00 | 2.6043E+02 | 3.4023E+02 | 5.1252E+02 |
| 1.8008E+00 | 2.8617E-02 | 4.8666E-02 | 6.4932E-02 | 3.5634E+00 | 2.7657E+02 | 3.4831E+02 | 4.8002E+02 | 5.1595E+00 | 2.9156E+02 | 3.8102E+02 | 5.7412E+02 |

```
1.8100E+00 2.6763E-02 4.5368E-02 5.9928E-02 3.5781E+00 2.8638E+02 3.6129E+02 5.0026E+02 5.1642E+00 3.2034E+02 4.1875E+02 6.3115E+02
1.8237E+00 2.3482E-02 3.9545E-02 5.2060E-02 3.5929E+00 2.8753E+02 3.6330E+02 5.0521E+02 5.1682E+00 3.4519E+02 4.5139E+02 6.8055E+02
1.8413E+00 2.4424E-02 4.0614E-02 5.3259E-02 3.6024E+00 2.8423E+02 3.5957E+02 5.0164E+02 5.1781E+00 3.5957E+02 4.7045E+02 7.0961E+02
1.8590E+00 3.3688E-02 5.4954E-02 7.1658E-02 3.6106E+00 2.7073E+02 3.4282E+02 4.7955E+02 5.1847E+00 3.4603E+02 4.5298E+02 6.8357E+02
1.8825E+00 5.3320E-02 8.4623E-02 1.0947E-01 3.6187E+00 2.3686E+02 3.0019E+02 4.2094E+02 5.1954E+00 3.2944E+02 4.3156E+02 6.5164E+02
1.9048E+00 7.2716E-02 1.1177E-01 1.4329E-01 3.6255E+00 2.0414E+02 2.5893E+02 3.6385E+02 5.2055E+00 3.3366E+02 4.3739E+02 6.6085E+02
1.9320E+00 9.5222E-02 1.4131E-01 1.7940E-01 3.6310E+00 1.6431E+02 2.0858E+02 2.9372E+02 5.2138E+00 3.4914E+02 4.5798E+02 6.9234E+02
1.9602E+00 9.8210E-02 1.4138E-01 1.7803E-01 3.6419E+00 1.2666E+02 1.6096E+02 2.2731E+02 5.2208E+00 3.6640E+02 4.8091E+02 7.2738E+02
1.9814E+00 9.3455E-02 1.3210E-01 1.6554E-01 3.6528E+00 1.0681E+02 1.3588E+02 1.9249E+02 5.2296E+00 3.9177E+02 5.1453E+02 7.7865E+02
2.0096E+00 1.1107E-01 1.5570E-01 1.9474E-01 3.6625E+00 9.6982E+01 1.2352E+02 1.7551E+02 5.2354E+00 4.0550E+02 5.3286E+02 8.0679E+02
2.0308E+00 1.3972E-01 1.9510E-01 2.4390E-01 3.6761E+00 9.4802E+01 1.2090E+02 1.7237E+02 5.2433E+00 4.0756E+02 5.3590E+02 8.1185E+02
2.0484E+00 1.6900E-01 2.3510E-01 2.9388E-01 3.6881E+00 9.8088E+01 1.2523E+02 1.7912E+02 5.2517E+00 3.9166E+02 5.1534E+02 7.8116E+02
2.0672E+00 1.9493E-01 2.6980E-01 3.3725E-01 3.6970E+00 1.0137E+02 1.2955E+02 1.8579E+02 5.2587E+00 3.5831E+02 4.7180E+02 7.1558E+02
2.0860E+00 2.1601E-01 2.9700E-01 3.7125E-01 3.7068E+00 1.0844E+02 1.3872E+02 1.9944E+02 5.2683E+00 3.3155E+02 4.3692E+02 6.6315E+02
2.1084E+00 2.1792E-01 2.9700E-01 3.7125E-01 3.7164E+00 1.1490E+02 1.4713E+02 2.1206E+02 5.2789E+00 3.2360E+02 4.2684E+02 6.4839E+02
2.1331E+00 2.5609E-01 3.4540E-01 4.3175E-01 3.7263E+00 1.1599E+02 1.4866E+02 2.1480E+02 5.2903E+00 3.3091E+02 4.3692E+02 6.6427E+02
2.1530E+00 3.2676E-01 4.3630E-01 5.4538E-01 3.7374E+00 1.0935E+02 1.4027E+02 2.0320E+02 5.3008E+00 3.4323E+02 4.5364E+02 6.9029E+02
2.1695E+00 4.5231E-01 5.9840E-01 7.4800E-01 3.7459E+00 1.0309E+02 1.3237E+02 1.9219E+02 5.3092E+00 3.5956E+02 4.7569E+02 7.2447E+02
2.1848E+00 6.2573E-01 8.2070E-01 1.0259E+00 3.7568E+00 9.9602E+01 1.2801E+02 1.8635E+02 5.3234E+00 3.8106E+02 5.0478E+02 7.6962E+02
2.2001E+00 8.4183E-01 1.0952E+00 1.3690E+00 3.7714E+00 1.0230E+02 1.3163E+02 1.9218E+02 5.3377E+00 3.9959E+02 5.3005E+02 8.0911E+02
2.2105E+00 1.0617E+00 1.3721E+00 1.7151E+00 3.7854E+00 1.0491E+02 1.3513E+02 1.9784E+02 5.3513E+00 3.9875E+02 5.2966E+02 8.0948E+02
2.2210E+00 1.2955E+00 1.6641E+00 2.0801E+00 3.7955E+00 1.0821E+02 1.3950E+02 2.0468E+02 5.3630E+00 3.7824E+02 5.0308E+02 7.6973E+02
2.2304E+00 1.5187E+00 1.9397E+00 2.4246E+00 3.8041E+00 1.0987E+02 1.4173E+02 2.0835E+02 5.3769E+00 3.5212E+02 4.6900E+02 7.1847E+02
2.2410E+00 1.7266E+00 2.1924E+00 2.7405E+00 3.8155E+00 1.1217E+02 1.4481E+02 2.1327E+02 5.3905E+00 3.4121E+02 4.5513E+02 6.9810E+02
2.2516E+00 1.8633E+00 2.3523E+00 2.9404E+00 3.8226E+00 1.1509E+02 1.4866E+02 2.1928E+02 5.4047E+00 3.5998E+02 4.8086E+02 7.3851E+02
2.2619E+00 2.0679E+00 2.5966E+00 3.2458E+00 3.8312E+00 1.1742E+02 1.5176E+02 2.2420E+02 5.4164E+00 3.9804E+02 5.3242E+02 8.1865E+02
2.2691E+00 2.2740E+00 2.8427E+00 3.5534E+00 3.8414E+00 1.2057E+02 1.5592E+02 2.3069E+02 5.4294E+00 4.4348E+02 5.9404E+02 9.1453E+02
2.2783E+00 2.5010E+00 3.1120E+00 3.8900E+00 3.8486E+00 1.2570E+02 1.6263E+02 2.4093E+02 5.4420E+00 4.6787E+02 6.2764E+02 9.6748E+02
2.2868E+00 2.8267E+00 3.5009E+00 4.3761E+00 3.8569E+00 1.2945E+02 1.6757E+02 2.4855E+02 5.4561E+00 5.0167E+02 6.7410E+02 1.0000E+03
2.2996E+00 2.8873E+00 3.5573E+00 4.4466E+00 3.8644E+00 1.3308E+02 1.7235E+02 2.5596E+02 5.4776E+00 5.7959E+02 7.8040E+02 1.0000E+03
2.3093E+00 2.9699E+00 3.6425E+00 4.5531E+00 3.8760E+00 1.3514E+02 1.7511E+02 2.6044E+02 5.4947E+00 6.1694E+02 8.3219E+02 1.0000E+03
2.3187E+00 3.1374E+00 3.8326E+00 4.7908E+00 3.8876E+00 1.3723E+02 1.7792E+02 2.6498E+02 5.5052E+00 5.7136E+02 7.7172E+02 1.0000E+03
2.3284E+00 3.3803E+00 4.1146E+00 5.1433E+00 3.8980E+00 1.4240E+02 1.8469E+02 2.7538E+02 5.5136E+00 4.9816E+02 6.7363E+02 1.0000E+03
2.3367E+00 3.6252E+00 4.3991E+00 5.4989E+00 3.9039E+00 1.5002E+02 1.9464E+02 2.9046E+02 5.5271E+00 4.4811E+02 6.0678E+02 9.4299E+02
2.3466E+00 3.8118E+00 4.6124E+00 5.7655E+00 3.9141E+00 1.6328E+02 2.1192E+02 3.1653E+02 5.5410E+00 4.6541E+02 6.3111E+02 9.8203E+02
2.3553E+00 3.7316E+00 4.5045E+00 5.6306E+00 3.9214E+00 1.7660E+02 2.2927E+02 3.4270E+02 5.5519E+00 4.9859E+02 6.7697E+02 1.0000E+03
2.3637E+00 3.5810E+00 4.3141E+00 5.3926E+00 3.9302E+00 1.9531E+02 2.5363E+02 3.7938E+02 5.5651E+00 5.1011E+02 6.9348E+02 1.0000E+03
2.3745E+00 3.6299E+00 4.3654E+00 5.4568E+00 3.9391E+00 2.1704E+02 2.8192E+02 4.2197E+02 5.5729E+00 4.7243E+02 6.4285E+02 1.0000E+03
2.3845E+00 3.7635E+00 4.5205E+00 5.6506E+00 3.9480E+00 2.4119E+02 3.1336E+02 4.6930E+02 5.5782E+00 4.2846E+02 5.8341E+02 9.1107E+02
2.3972E+00 3.7801E+00 4.5374E+00 5.6718E+00 3.9585E+00 2.7958E+02 3.6330E+02 5.4437E+02 5.5839E+00 3.7967E+02 5.1732E+02 8.0833E+02
2.4031E+00 3.2776E+00 3.9339E+00 4.9174E+00 3.9674E+00 3.2075E+02 4.1687E+02 6.2489E+02 5.5912E+00 3.3024E+02 4.5032E+02 7.0413E+02
2.4120E+00 2.3857E+00 2.8646E+00 3.5808E+00 3.9764E+00 3.6220E+02 4.7079E+02 7.0592E+02 5.5993E+00 2.9970E+02 4.0905E+02 6.4012E+02
2.4192E+00 1.8298E+00 2.1992E+00 2.7490E+00 3.9838E+00 4.0033E+02 5.2039E+02 7.8044E+02 5.6121E+00 2.8073E+02 3.8363E+02 6.0097E+02
2.4348E+00 1.5497E+00 1.8663E+00 2.3329E+00 3.9942E+00 4.3585E+02 5.6659E+02 8.4984E+02 5.6242E+00 2.8804E+02 3.9412E+02 6.1807E+02
2.4480E+00 1.6322E+00 1.9707E+00 2.4634E+00 4.0048E+00 4.5932E+02 5.9712E+02 8.9567E+02 5.6379E+00 3.1806E+02 4.3571E+02 6.8403E+02
2.4596E+00 1.7867E+00 2.1632E+00 2.7040E+00 4.0138E+00 4.8677E+02 6.3280E+02 9.4919E+02 5.6452E+00 3.5199E+02 4.8261E+02 7.5821E+02
2.4700E+00 1.8998E+00 2.3068E+00 2.8835E+00 4.0245E+00 5.1340E+02 6.6742E+02 1.0000E+03 5.6516E+00 3.9121E+02 5.3673E+02 8.4372E+02
2.4822E+00 2.0490E+00 2.4957E+00 3.1196E+00 4.0425E+00 5.1340E+02 6.6742E+02 1.0000E+03 5.6562E+00 4.2596E+02 5.8470E+02 9.1953E+02
2.4889E+00 2.1429E+00 2.6167E+00 3.2709E+00 4.0624E+00 5.0530E+02 6.5689E+02 9.8533E+02 5.6607E+00 4.6848E+02 6.4338E+02 1.0000E+03
2.4986E+00 2.2402E+00 2.7435E+00 3.4294E+00 4.0730E+00 4.9460E+02 6.4298E+02 9.6447E+02 5.6671E+00 5.3968E+02 7.4157E+02 1.0000E+03
2.5100E+00 2.4246E+00 2.9787E+00 3.7234E+00 4.0854E+00 4.8948E+02 6.3633E+02 9.5449E+02 5.6720E+00 6.0503E+02 8.3175E+02 1.0000E+03
2.5193E+00 2.5152E+00 3.0992E+00 3.8740E+00 4.0960E+00 4.8677E+02 6.3280E+02 9.4920E+02 5.6763E+00 6.8668E+02 9.4437E+02 1.0000E+03
2.5300E+00 2.6608E+00 3.2881E+00 4.1101E+00 4.1053E+00 4.7681E+02 6.1985E+02 9.2978E+02 5.6784E+00 7.1880E+02 9.8883E+02 1.0000E+03
2.5402E+00 2.8254E+00 3.5009E+00 4.3761E+00 4.1145E+00 4.6409E+02 6.0332E+02 9.0498E+02 5.6814E+00 7.7750E+02 1.0000E+03 1.0000E+03
2.5494E+00 2.9454E+00 3.6581E+00 4.5726E+00 4.1208E+00 4.4992E+02 5.8490E+02 8.7735E+02 5.6885E+00 9.0768E+02 1.0000E+03 1.0000E+03
2.5597E+00 3.1330E+00 3.8993E+00 4.8741E+00 4.1306E+00 4.4165E+02 5.7415E+02 8.6122E+02 5.7000E+00 1.0000E+03 1.0000E+03 1.0000E+03
2.5712E+00 3.3915E+00 4.2287E+00 5.2859E+00 4.1392E+00 4.5714E+02 5.9428E+02 8.9141E+02 5.7126E+00 1.0000E+03 1.0000E+03 1.0000E+03
2.5809E+00 3.6344E+00 4.5374E+00 5.6717E+00 4.1472E+00 4.7909E+02 6.2282E+02 9.3422E+02 5.7210E+00 1.0000E+03 1.0000E+03 1.0000E+03
2.5886E+00 3.8160E+00 4.7676E+00 5.9595E+00 4.1533E+00 5.0812E+02 6.6056E+02 9.9083E+02 5.7700E+00 1.0000E+03 1.0000E+03 1.0000E+03
2.5988E+00 4.1426E+00 5.1775E+00 6.4717E+00 4.1596E+00 5.3252E+02 6.9228E+02 1.0000E+03 5.7818E+00 9.3692E+02 1.0000E+03 1.0000E+03
2.6100E+00 4.3860E+00 5.4816E+00 6.8511E+00 4.1674E+00 5.6166E+02 7.3015E+02 1.0000E+03 5.7870E+00 8.5000E+02 1.0000E+03 1.0000E+03
2.6202E+00 4.5063E+00 5.6308E+00 7.0359E+00 4.1784E+00 5.6749E+02 7.3774E+02 1.0000E+03 5.7911E+00 7.7051E+02 1.0000E+03 1.0000E+03
2.6328E+00 4.7423E+00 5.9235E+00 7.3989E+00 4.1879E+00 5.4106E+02 7.0337E+02 1.0000E+03 5.7940E+00 7.1994E+02 1.0000E+03 1.0000E+03
2.6395E+00 4.9080E+00 6.1281E+00 7.6516E+00 4.2035E+00 5.2163E+02 6.7811E+02 1.0000E+03 5.7961E+00 6.8831E+02 9.5629E+02 1.0000E+03
2.6462E+00 5.1996E+00 6.4893E+00 8.0989E+00 4.2256E+00 5.4106E+02 7.0337E+02 1.0000E+03 5.8005E+00 6.0464E+02 8.4040E+02 1.0000E+03
2.6563E+00 5.6023E+00 6.9877E+00 8.7156E+00 4.2337E+00 5.5854E+02 7.2610E+02 1.0000E+03 5.8136E+00 4.4000E+02 6.1209E+02 9.7390E+02
2.6649E+00 5.9186E+00 7.3772E+00 9.1952E+00 4.2433E+00 5.8258E+02 7.5735E+02 1.0000E+03 5.8350E+00 2.8686E+02 3.9954E+02 6.3640E+02
2.6738E+00 5.9836E+00 7.4523E+00 9.2816E+00 4.2495E+00 6.0140E+02 7.8182E+02 1.0000E+03 5.8621E+00 2.2484E+02 3.1360E+02 5.0013E+02
2.6824E+00 5.6528E+00 7.0339E+00 8.7524E+00 4.2575E+00 6.3430E+02 8.2459E+02 1.0000E+03 5.8921E+00 2.0016E+02 2.7956E+02 4.4637E+02
2.6956E+00 4.7037E+00 5.8460E+00 7.2657E+00 4.2656E+00 6.5116E+02 8.4651E+02 1.0000E+03 5.9327E+00 2.0486E+02 2.8647E+02 4.5789E+02
2.7056E+00 4.3423E+00 5.3906E+00 6.6920E+00 4.2719E+00 6.4089E+02 8.3316E+02 1.0000E+03 5.9656E+00 2.2016E+02 3.0810E+02 4.9278E+02
2.7141E+00 4.2169E+00 5.2292E+00 6.4845E+00 4.2801E+00 6.0813E+02 7.9057E+02 1.0000E+03 6.0280E+00 2.5887E+02 3.6238E+02 5.7977E+02
2.7229E+00 4.2650E+00 5.2825E+00 6.5427E+00 4.2849E+00 5.7384E+02 7.4599E+02 1.0000E+03 6.0769E+00 3.1031E+02 4.3443E+02 6.9508E+02
2.7354E+00 4.5854E+00 5.6709E+00 7.0133E+00 4.2912E+00 5.3549E+02 6.9614E+02 1.0000E+03 6.1395E+00 4.1120E+02 5.7568E+02 9.2108E+02
2.7464E+00 4.9273E+00 6.0848E+00 7.5142E+00 4.2978E+00 4.9222E+02 6.3988E+02 9.5982E+02 6.1600E+00 4.5786E+02 6.4100E+02 1.0000E+03
2.7553E+00 5.2088E+00 6.4239E+00 7.9224E+00 4.3013E+00 4.6018E+02 5.9824E+02 8.9736E+02 6.1823E+00 5.2017E+02 7.2824E+02 1.0000E+03
2.7636E+00 5.4957E+00 6.7689E+00 8.3370E+00 4.3067E+00 4.3250E+02 5.6225E+02 8.4337E+02 6.2000E+00 5.8714E+02 8.2200E+02 1.0000E+03
2.7733E+00 5.9015E+00 7.2582E+00 8.9267E+00 4.3124E+00 4.0898E+02 5.3168E+02 7.9752E+02 6.2125E+00 6.4071E+02 8.9700E+02 1.0000E+03
2.7843E+00 6.2757E+00 7.7058E+00 9.4618E+00 4.3221E+00 3.7984E+02 4.9379E+02 7.4068E+02 6.2214E+00 6.8356E+02 9.5699E+02 1.0000E+03
2.7959E+00 6.6054E+00 8.0966E+00 9.9244E+00 4.3351E+00 3.8592E+02 5.0170E+02 7.5255E+02 6.2267E+00 7.1071E+02 9.9500E+02 1.0000E+03
2.8077E+00 6.9188E+00 8.4656E+00 1.0358E+01 4.3480E+00 3.9839E+02 5.1791E+02 7.7687E+02 6.2334E+00 7.5000E+02 1.0000E+03 1.0000E+03
2.8197E+00 7.3600E+00 8.9894E+00 1.0980E+01 4.3596E+00 3.7773E+02 4.9105E+02 7.3657E+02 6.2498E+00 8.4286E+02 1.0000E+03 1.0000E+03
2.8301E+00 7.4042E+00 9.0281E+00 1.1008E+01 4.3692E+00 3.6013E+02 4.6817E+02 7.0226E+02 6.2650E+00 9.5000E+02 1.0000E+03 1.0000E+03
2.8432E+00 7.2240E+00 8.7930E+00 1.0703E+01 4.3834E+00 3.8410E+02 4.9933E+02 7.4900E+02 6.2800E+00 1.0000E+03 1.0000E+03 1.0000E+03
2.8530E+00 7.2433E+00 8.8032E+00 1.0699E+01 4.3926E+00 4.1092E+02 5.3419E+02 8.0129E+02 6.3103E+00 1.0000E+03 1.0000E+03 1.0000E+03
2.8621E+00 7.3360E+00 8.9037E+00 1.0806E+01 4.4018E+00 4.0240E+02 5.2312E+02 7.8468E+02 6.3500E+00 1.0000E+03 1.0000E+03 1.0000E+03
```

## Section 2.

MATLAB® function for calculating refractive indices and absorption coefficient. The function was written and tested using MATLAB Version 9.10 (R2021a). Users can make their own modifications.

```matlab
function ValueToReturn=AbsorptionBBO_v0(Wavelength_mkm)
% Function returns minimal typical and maximal values of attenuation
%    as power loss 'dB/cm', and refractive indices of o and e polarizations
% T=10^(-dB/10*L) - transmittance
%
% INPUT is a vector wavelength in mkm
% OUTPUT=[wavelength; min; typ; max; o; e]
% Attenuation values are clipped at 150 dB/cm in UV, 1000 dB/cm in IR
%
% Attenuation coefficient is valid 0.188-6.22 mkm
% Sellmeier equation valid 0.188-5.200 mkm, Opt.Mater.Express,8,1410-1418(2018)
%
% Date: 2021.11.01
%
% Example:
% SomeData=AbsorptionBBO_v0([0.532; 1.064; 1.550]);

%checking input is valid
if exist('Wavelength_mkm', 'var')==0 || isempty(Wavelength_mkm)==1 || isnumeric(Wavelength_mkm)~=1
    disp('Warning, input is empty or not a number in function ''AbsorptionBBO_v0''');
    ValueToReturn=[];
    return
end

inputSize=size(Wavelength_mkm);
if inputSize(2)<inputSize(1) && min(inputSize)==1
    inputValues=Wavelength_mkm;
elseif min(inputSize)==1
    inputValues=Wavelength_mkm.';
else
    disp('Warning, input must be a vector in function ''AbsorptionBBO_v0''');
    ValueToReturn=[];
    return
end % end checking input is valid

persistent TableOfBBOabsorptionCoef_iwfjp
    %global TableOfBBOabsorptionCoef_iwfjp %if other function are using it
%creates variable if not yet exists; otherwise uses existing persistent
if isempty(TableOfBBOabsorptionCoef_iwfjp)
    disp('persistent variable is created in function ''AbsorptionBBO_v0''');
    TableOfBBOabsorptionCoef_iwfjp=[...
1.8500E-01 1.5000E+02 1.5000E+02 1.5000E+02; 1.8530E-01 1.5000E+02 1.5000E+02 1.5000E+02; 1.8590E-01 1.5000E+02 1.5000E+02 1.5000E+02;...
1.8640E-01 1.3157E+02 1.5000E+02 1.5000E+02; 1.8679E-01 1.1200E+02 1.5000E+02 1.5000E+02; 1.8701E-01 1.0324E+02 1.5000E+02 1.5000E+02;...
1.8745E-01 8.6507E+01 1.5000E+02 1.5000E+02; 1.8810E-01 6.7513E+01 1.2760E+02 1.5000E+02; 1.8870E-01 5.4510E+01 1.0329E+02 1.5000E+02;...
1.8930E-01 4.3283E+01 8.2200E+01 1.5000E+02; 1.9021E-01 3.0908E+01 5.8907E+01 1.5000E+02; 1.9100E-01 2.2154E+01 4.2600E+01 1.2808E+02;...
1.9150E-01 1.7920E+01 3.5000E+01 1.0553E+02; 1.9204E-01 1.3984E+01 2.8198E+01 8.5532E+01; 1.9399E-01 6.2526E+00 1.3500E+01 4.1499E+01;...
1.9597E-01 2.8122E+00 6.5564E+00 2.0477E+01; 1.9746E-01 1.5182E+00 3.7480E+00 1.1854E+01; 1.9913E-01 8.0279E-01 2.0557E+00 6.5599E+00;...
2.0059E-01 4.7717E-01 1.2426E+00 3.9932E+00; 2.0241E-01 2.8270E-01 7.4901E-01 2.4441E+00; 2.0459E-01 1.7431E-01 4.7702E-01 1.6080E+00;...
2.0748E-01 1.0629E-01 3.0757E-01 1.0975E+00; 2.1239E-01 6.9493E-02 2.1715E-01 8.4060E-01; 2.1978E-01 4.6592E-02 1.5449E-01 6.4696E-01;...
2.2664E-01 3.3634E-02 1.1258E-01 5.0944E-01; 2.3555E-01 2.3632E-02 7.8577E-02 4.0108E-01; 2.4777E-01 1.4629E-02 4.9864E-02 2.9194E-01;...
2.6379E-01 7.9651E-03 3.1086E-02 2.1196E-01; 2.8009E-01 4.3116E-03 1.9992E-02 1.6043E-01; 2.9662E-01 2.4107E-03 1.3427E-02 1.2846E-01;...
3.1798E-01 1.3960E-03 8.4544E-03 8.8707E-02; 3.3000E-01 1.0057E-03 6.3708E-03 6.8134E-02; 3.4702E-01 6.8104E-04 4.5391E-03 4.6718E-02;...
3.6000E-01 5.2113E-04 3.5478E-03 3.3776E-02; 3.7556E-01 3.8512E-04 2.6533E-03 2.2261E-02; 3.9000E-01 3.0126E-04 2.0939E-03 1.6431E-02;...
4.1156E-01 2.2589E-04 1.5837E-03 1.2194E-02; 4.3000E-01 1.8176E-04 1.2859E-03 1.0158E-02; 4.5088E-01 1.4566E-04 1.0425E-03 8.6360E-03;...
4.7500E-01 1.1791E-04 8.5507E-04 7.4511E-03; 4.9996E-01 9.8396E-05 7.2497E-04 6.5879E-03; 5.4934E-01 7.1447E-05 5.4365E-04 5.2204E-03;...
5.9999E-01 5.5989E-05 4.4213E-04 4.4877E-03; 6.4824E-01 4.4599E-05 3.6501E-04 3.8705E-03; 7.0098E-01 3.6078E-05 3.0532E-04 3.3459E-03;...
7.5036E-01 3.0249E-05 2.6250E-04 2.8987E-03; 7.9973E-01 2.6049E-05 2.3197E-04 2.5586E-03; 8.4462E-01 2.3588E-05 2.1362E-04 2.3516E-03;...
8.9624E-01 2.2470E-05 2.0499E-04 2.2551E-03; 9.5571E-01 2.2184E-05 2.0358E-04 2.2378E-03; 1.0017E+00 2.1964E-05 2.0280E-04 2.2221E-03;...
1.0350E+00 2.1884E-05 2.0280E-04 2.2007E-03; 1.0714E+00 2.2734E-05 2.0290E-04 2.1844E-03; 1.1651E+00 2.7907E-05 2.0384E-04 2.2298E-03;...
1.2000E+00 3.3916E-05 2.0511E-04 2.2492E-03; 1.2426E+00 4.0699E-05 2.0689E-04 2.1689E-03; 1.3130E+00 5.8122E-05 2.3855E-04 2.2640E-03;...
1.3735E+00 9.2663E-05 3.1658E-04 2.5361E-03; 1.4233E+00 1.4307E-04 4.3615E-04 3.0670E-03; 1.4786E+00 2.6365E-04 7.1681E-04 4.3400E-03;...
1.5218E+00 4.6930E-04 1.1256E-03 5.7321E-03; 1.5681E+00 8.6597E-04 1.8544E-03 7.9829E-03; 1.6139E+00 1.5595E-03 3.1450E-03 1.2130E-02;...
1.6517E+00 2.5993E-03 5.0696E-03 1.7459E-02; 1.6798E+00 3.9662E-03 7.5127E-03 2.2440E-02; 1.7056E+00 6.2740E-03 1.1536E-02 2.8993E-02;...
1.7318E+00 1.0280E-02 1.8379E-02 3.8158E-02; 1.7493E+00 1.5030E-02 2.6322E-02 4.6357E-02; 1.7687E+00 2.2052E-02 3.8021E-02 5.8097E-02;...
1.7878E+00 2.8513E-02 4.8691E-02 6.7548E-02; 1.8008E+00 2.8617E-02 4.8666E-02 6.4932E-02; 1.8100E+00 2.6763E-02 4.5368E-02 5.9928E-02;...
1.8237E+00 2.3482E-02 3.9545E-02 5.2060E-02; 1.8413E+00 2.4424E-02 4.0614E-02 5.3259E-02; 1.8590E+00 3.3688E-02 5.4954E-02 7.1658E-02;...
1.8825E+00 5.3320E-02 8.4623E-02 1.0947E-01; 1.9048E+00 7.2716E-02 1.1178E-01 1.4329E-01; 1.9320E+00 9.5222E-02 1.4131E-01 1.7940E-01;...
1.9602E+00 9.8210E-02 1.4138E-01 1.7803E-01; 1.9814E+00 9.3455E-02 1.3210E-01 1.6554E-01; 2.0096E+00 1.1107E-01 1.5570E-01 1.9474E-01;...
2.0308E+00 1.3972E-01 1.9510E-01 2.4390E-01; 2.0484E+00 1.6900E-01 2.3510E-01 2.9388E-01; 2.0672E+00 1.9493E-01 2.6980E-01 3.3725E-01;...
```

```
2.0860E+00 2.1601E-01 2.9700E-01 3.7125E-01; 2.1084E+00 2.1792E-01 2.9700E-01 3.7125E-01; 2.1331E+00 2.5609E-01 3.4540E-01 4.3175E-01;...
2.1530E+00 3.2676E-01 4.3630E-01 5.4538E-01; 2.1695E+00 4.5231E-01 5.9840E-01 7.4800E-01; 2.1848E+00 6.2573E-01 8.2070E-01 1.0259E+00;...
2.2001E+00 8.4183E-01 1.0952E+00 1.3690E+00; 2.2105E+00 1.0617E+00 1.3721E+00 1.7151E+00; 2.2210E+00 1.2955E+00 1.6641E+00 2.0801E+00;...
2.2304E+00 1.5187E+00 1.9397E+00 2.4246E+00; 2.2410E+00 1.7266E+00 2.1924E+00 2.7405E+00; 2.2516E+00 1.8633E+00 2.3523E+00 2.9404E+00;...
2.2619E+00 2.0679E+00 2.5966E+00 3.2458E+00; 2.2691E+00 2.2740E+00 2.8427E+00 3.5534E+00; 2.2783E+00 2.5010E+00 3.1120E+00 3.8900E+00;...
2.2868E+00 2.8267E+00 3.5009E+00 4.3761E+00; 2.2996E+00 2.8873E+00 3.5573E+00 4.4466E+00; 2.3093E+00 2.9699E+00 3.6425E+00 4.5531E+00;...
2.3187E+00 3.1374E+00 3.8326E+00 4.7908E+00; 2.3284E+00 3.3803E+00 4.1146E+00 5.1433E+00; 2.3367E+00 3.6252E+00 4.3991E+00 5.4989E+00;...
2.3466E+00 3.8118E+00 4.6124E+00 5.7655E+00; 2.3553E+00 3.7316E+00 4.5045E+00 5.6306E+00; 2.3637E+00 3.5810E+00 4.3141E+00 5.3926E+00;...
2.3745E+00 3.6299E+00 4.3654E+00 5.4568E+00; 2.3845E+00 3.7635E+00 4.5205E+00 5.6506E+00; 2.3972E+00 3.7801E+00 4.5374E+00 5.6718E+00;...
2.4031E+00 3.2776E+00 3.9339E+00 4.9174E+00; 2.4120E+00 2.3857E+00 2.8646E+00 3.5808E+00; 2.4192E+00 1.8298E+00 2.1992E+00 2.7490E+00;...
2.4348E+00 1.5497E+00 1.8663E+00 2.3329E+00; 2.4480E+00 1.6322E+00 1.9707E+00 2.4634E+00; 2.4596E+00 1.7867E+00 2.1632E+00 2.7040E+00;...
2.4700E+00 1.8998E+00 2.3068E+00 2.8835E+00; 2.4822E+00 2.0490E+00 2.4957E+00 3.1196E+00; 2.4889E+00 2.1429E+00 2.6167E+00 3.2709E+00;...
2.4986E+00 2.2402E+00 2.7435E+00 3.4294E+00; 2.5100E+00 2.4246E+00 2.9787E+00 3.7234E+00; 2.5193E+00 2.5152E+00 3.0992E+00 3.8740E+00;...
2.5300E+00 2.6608E+00 3.2881E+00 4.1101E+00; 2.5402E+00 2.8254E+00 3.5009E+00 4.3761E+00; 2.5494E+00 2.9454E+00 3.6581E+00 4.5726E+00;...
2.5597E+00 3.1330E+00 3.8993E+00 4.8741E+00; 2.5712E+00 3.3915E+00 4.2287E+00 5.2859E+00; 2.5809E+00 3.6344E+00 4.5374E+00 5.6717E+00;...
2.5886E+00 3.8160E+00 4.7676E+00 5.9595E+00; 2.5988E+00 4.1426E+00 5.1775E+00 6.4717E+00; 2.6100E+00 4.3860E+00 5.4816E+00 6.8511E+00;...
2.6202E+00 4.5063E+00 5.6308E+00 7.0359E+00; 2.6328E+00 4.7423E+00 5.9235E+00 7.3989E+00; 2.6395E+00 4.9080E+00 6.1281E+00 7.6516E+00;...
2.6462E+00 5.1996E+00 6.4893E+00 8.0989E+00; 2.6563E+00 5.6023E+00 6.9877E+00 8.7156E+00; 2.6649E+00 5.9186E+00 7.3772E+00 9.1952E+00;...
2.6738E+00 5.9836E+00 7.4523E+00 9.2816E+00; 2.6824E+00 5.6528E+00 7.0339E+00 8.7524E+00; 2.6956E+00 4.7037E+00 5.8460E+00 7.2657E+00;...
2.7056E+00 4.3423E+00 5.3906E+00 6.6920E+00; 2.7141E+00 4.2169E+00 5.2292E+00 6.4845E+00; 2.7229E+00 4.2650E+00 5.2825E+00 6.5427E+00;...
2.7354E+00 4.5854E+00 5.6709E+00 7.0133E+00; 2.7464E+00 4.9273E+00 6.0848E+00 7.5142E+00; 2.7553E+00 5.2088E+00 6.4239E+00 7.9224E+00;...
2.7636E+00 5.4957E+00 6.7689E+00 8.3370E+00; 2.7733E+00 5.9015E+00 7.2582E+00 8.9267E+00; 2.7843E+00 6.2757E+00 7.7058E+00 9.4618E+00;...
2.7959E+00 6.6054E+00 8.0966E+00 9.9244E+00; 2.8077E+00 6.9188E+00 8.4656E+00 1.0358E+01; 2.8197E+00 7.3600E+00 8.9894E+00 1.0980E+01;...
2.8301E+00 7.4042E+00 9.0281E+00 1.1008E+01; 2.8432E+00 7.2240E+00 8.7930E+00 1.0703E+01; 2.8530E+00 7.2433E+00 8.8032E+00 1.0699E+01;...
2.8621E+00 7.3360E+00 8.9037E+00 1.0806E+01; 2.8728E+00 7.6809E+00 9.3095E+00 1.1283E+01; 2.8824E+00 8.2230E+00 9.9542E+00 1.2050E+01;...
2.8901E+00 8.7495E+00 1.0580E+01 1.2794E+01; 2.8992E+00 9.1253E+00 1.1022E+01 1.3313E+01; 2.9101E+00 9.6453E+00 1.1637E+01 1.4039E+01;...
2.9212E+00 1.0331E+01 1.2450E+01 1.5004E+01; 2.9329E+00 1.0774E+01 1.2970E+01 1.5615E+01; 2.9421E+00 1.0783E+01 1.2970E+01 1.5602E+01;...
2.9520E+00 1.0507E+01 1.2630E+01 1.5182E+01; 2.9602E+00 9.8323E+00 1.1813E+01 1.4191E+01; 2.9725E+00 9.9276E+00 1.1921E+01 1.4315E+01;...
2.9795E+00 1.0446E+01 1.2540E+01 1.5053E+01; 2.9907E+00 1.1112E+01 1.3336E+01 1.6005E+01; 2.9998E+00 1.1113E+01 1.3336E+01 1.6004E+01;...
3.0093E+00 1.0822E+01 1.2987E+01 1.5586E+01; 3.0184E+00 1.1730E+01 1.4077E+01 1.6898E+01; 3.0273E+00 1.2639E+01 1.5170E+01 1.8215E+01;...
3.0369E+00 1.3715E+01 1.6464E+01 1.9778E+01; 3.0484E+00 1.4776E+01 1.7742E+01 2.1328E+01; 3.0598E+00 1.5522E+01 1.8643E+01 2.2430E+01;...
3.0709E+00 1.6242E+01 1.9514E+01 2.3500E+01; 3.0801E+00 1.5682E+01 1.8847E+01 2.2721E+01; 3.0895E+00 1.5373E+01 1.8483E+01 2.2307E+01;...
3.1000E+00 1.6009E+01 1.9256E+01 2.3267E+01; 3.1080E+00 1.5942E+01 1.9182E+01 2.3204E+01; 3.1139E+00 1.4885E+01 1.7916E+01 2.1697E+01;...
3.1237E+00 1.3412E+01 1.6151E+01 1.9587E+01; 3.1327E+00 1.2759E+01 1.5373E+01 1.8673E+01; 3.1445E+00 1.3656E+01 1.6464E+01 2.0037E+01;...
3.1550E+00 1.4884E+01 1.7958E+01 2.1900E+01; 3.1670E+00 1.5624E+01 1.8865E+01 2.3060E+01; 3.1790E+00 1.6482E+01 1.9917E+01 2.4403E+01;...
3.1878E+00 1.7514E+01 2.1179E+01 2.6003E+01; 3.1964E+00 1.9249E+01 2.3294E+01 2.8660E+01; 3.2068E+00 2.1564E+01 2.6117E+01 3.2211E+01;...
3.2176E+00 2.4681E+01 2.9919E+01 3.7002E+01; 3.2316E+00 2.9401E+01 3.5680E+01 4.4268E+01; 3.2435E+00 3.5235E+01 4.2805E+01 5.3272E+01;...
3.2540E+00 4.1919E+01 5.0973E+01 6.3613E+01; 3.2615E+00 4.8339E+01 5.8827E+01 7.3590E+01; 3.2704E+00 5.6486E+01 6.8802E+01 8.6282E+01;...
3.2801E+00 6.4664E+01 7.8836E+01 9.9135E+01; 3.2896E+00 7.6027E+01 9.2780E+01 1.1700E+02; 3.3007E+00 8.7526E+01 1.0693E+02 1.3525E+02;...
3.3098E+00 9.5709E+01 1.1704E+02 1.4847E+02; 3.3196E+00 1.0303E+02 1.2612E+02 1.6045E+02; 3.3279E+00 1.1013E+02 1.3493E+02 1.7215E+02;...
3.3377E+00 1.2529E+02 1.5367E+02 1.9663E+02; 3.3467E+00 1.4354E+02 1.7625E+02 2.2624E+02; 3.3595E+00 1.6512E+02 2.0302E+02 2.6158E+02;...
3.3720E+00 1.8921E+02 2.3295E+02 3.0130E+02; 3.3821E+00 2.1581E+02 2.6602E+02 3.4532E+02; 3.3936E+00 2.3823E+02 2.9405E+02 3.8313E+02;...
3.4057E+00 2.6325E+02 3.2537E+02 4.2555E+02; 3.4166E+00 2.8597E+02 3.5389E+02 4.6450E+02; 3.4253E+00 2.9751E+02 3.6857E+02 4.8527E+02;...
3.4346E+00 3.0556E+02 3.7894E+02 5.0036E+02; 3.4410E+00 3.1018E+02 3.8501E+02 5.0964E+02; 3.4488E+00 3.2169E+02 3.9967E+02 5.3042E+02;...
3.4580E+00 3.3522E+02 4.1687E+02 5.5472E+02; 3.4630E+00 3.5696E+02 4.4424E+02 5.9241E+02; 3.4697E+00 3.8588E+02 4.8062E+02 6.4236E+02;...
3.4775E+00 4.0438E+02 5.0410E+02 6.7539E+02; 3.4840E+00 4.1511E+02 5.1791E+02 6.9555E+02; 3.4919E+00 3.9540E+02 4.9379E+02 6.6489E+02;...
3.5010E+00 3.4233E+02 4.2795E+02 5.7789E+02; 3.5089E+00 3.0595E+02 3.8287E+02 5.1853E+02; 3.5182E+00 2.8087E+02 3.5193E+02 4.7826E+02;...
3.5341E+00 2.7013E+02 3.3902E+02 4.6280E+02; 3.5461E+00 2.6841E+02 3.3741E+02 4.6266E+02; 3.5634E+00 2.7657E+02 3.4831E+02 4.8002E+02;...
3.5781E+00 2.8638E+02 3.6129E+02 5.0026E+02; 3.5929E+00 2.8753E+02 3.6330E+02 5.0521E+02; 3.6024E+00 2.8423E+02 3.5957E+02 5.0164E+02;...
3.6106E+00 2.7073E+02 3.4282E+02 4.7955E+02; 3.6187E+00 2.3686E+02 3.0019E+02 4.2094E+02; 3.6255E+00 2.0414E+02 2.5893E+02 3.6385E+02;...
3.6310E+00 1.6431E+02 2.0858E+02 2.9372E+02; 3.6419E+00 1.2666E+02 1.6096E+02 2.2731E+02; 3.6528E+00 1.0681E+02 1.3588E+02 1.9249E+02;...
3.6625E+00 9.6982E+01 1.2352E+02 1.7551E+02; 3.6761E+00 9.4802E+01 1.2090E+02 1.7237E+02; 3.6881E+00 9.8088E+01 1.2523E+02 1.7912E+02;...
3.6970E+00 1.0137E+02 1.2955E+02 1.8579E+02; 3.7068E+00 1.0844E+02 1.3872E+02 1.9944E+02; 3.7164E+00 1.1490E+02 1.4713E+02 2.1206E+02;...
3.7263E+00 1.1599E+02 1.4866E+02 2.1480E+02; 3.7374E+00 1.0935E+02 1.4027E+02 2.0320E+02; 3.7459E+00 1.0309E+02 1.3237E+02 1.9219E+02;...
3.7568E+00 9.9602E+01 1.2801E+02 1.8635E+02; 3.7714E+00 1.0230E+02 1.3163E+02 1.9218E+02; 3.7854E+00 1.0491E+02 1.3513E+02 1.9784E+02;...
3.7955E+00 1.0821E+02 1.3950E+02 2.0468E+02; 3.8041E+00 1.0987E+02 1.4173E+02 2.0835E+02; 3.8155E+00 1.1217E+02 1.4481E+02 2.1327E+02;...
3.8226E+00 1.1509E+02 1.4866E+02 2.1928E+02; 3.8312E+00 1.1742E+02 1.5176E+02 2.2420E+02; 3.8414E+00 1.2057E+02 1.5592E+02 2.3069E+02;...
3.8486E+00 1.2570E+02 1.6263E+02 2.4093E+02; 3.8569E+00 1.2945E+02 1.6757E+02 2.4855E+02; 3.8644E+00 1.3308E+02 1.7235E+02 2.5596E+02;...
3.8760E+00 1.3514E+02 1.7511E+02 2.6044E+02; 3.8876E+00 1.3723E+02 1.7792E+02 2.6498E+02; 3.8980E+00 1.4240E+02 1.8469E+02 2.7538E+02;...
3.9039E+00 1.5002E+02 1.9464E+02 2.9046E+02; 3.9141E+00 1.6328E+02 2.1192E+02 3.1653E+02; 3.9214E+00 1.7660E+02 2.2927E+02 3.4270E+02;...
3.9302E+00 1.9531E+02 2.5363E+02 3.7938E+02; 3.9391E+00 2.1704E+02 2.8192E+02 4.2197E+02; 3.9480E+00 2.4119E+02 3.1336E+02 4.6930E+02;...
3.9585E+00 2.7958E+02 3.6330E+02 5.4437E+02; 3.9674E+00 3.2075E+02 4.1687E+02 6.2489E+02; 3.9764E+00 3.6220E+02 4.7079E+02 7.0592E+02;...
3.9838E+00 4.0033E+02 5.2039E+02 7.8044E+02; 3.9942E+00 4.3585E+02 5.6659E+02 8.4984E+02; 4.0048E+00 4.5932E+02 5.9712E+02 8.9567E+02;...
4.0138E+00 4.8677E+02 6.3280E+02 9.4919E+02; 4.0245E+00 5.1340E+02 6.6742E+02 1.0000E+03; 4.0425E+00 5.1340E+02 6.6742E+02 1.0000E+03;...
4.0624E+00 5.0530E+02 6.5689E+02 9.8533E+02; 4.0730E+00 4.9460E+02 6.4298E+02 9.6447E+02; 4.0854E+00 4.8948E+02 6.3633E+02 9.5449E+02;...
4.0960E+00 4.8677E+02 6.3280E+02 9.4920E+02; 4.1053E+00 4.7681E+02 6.1985E+02 9.2978E+02; 4.1145E+00 4.6409E+02 6.0332E+02 9.0498E+02;...
4.1208E+00 4.4992E+02 5.8490E+02 8.7735E+02; 4.1306E+00 4.4165E+02 5.7415E+02 8.6122E+02; 4.1392E+00 4.5714E+02 5.9428E+02 8.9141E+02;...
4.1472E+00 4.7909E+02 6.2282E+02 9.3422E+02; 4.1533E+00 5.0812E+02 6.6056E+02 9.9083E+02; 4.1596E+00 5.3252E+02 6.9228E+02 1.0000E+03;...
4.1674E+00 5.6166E+02 7.3015E+02 1.0000E+03; 4.1784E+00 5.6749E+02 7.3774E+02 1.0000E+03; 4.1879E+00 5.4106E+02 7.0337E+02 1.0000E+03;...
4.2035E+00 5.2163E+02 6.7811E+02 1.0000E+03; 4.2256E+00 5.4106E+02 7.0337E+02 1.0000E+03; 4.2337E+00 5.5854E+02 7.2610E+02 1.0000E+03;...
4.2433E+00 5.8258E+02 7.5735E+02 1.0000E+03; 4.2495E+00 6.0140E+02 7.8182E+02 1.0000E+03; 4.2575E+00 6.3430E+02 8.2459E+02 1.0000E+03;...
4.2656E+00 6.5116E+02 8.4651E+02 1.0000E+03; 4.2719E+00 6.4089E+02 8.3316E+02 1.0000E+03; 4.2801E+00 6.0813E+02 7.9057E+02 1.0000E+03;...
4.2849E+00 5.7384E+02 7.4599E+02 1.0000E+03; 4.2912E+00 5.3549E+02 6.9614E+02 1.0000E+03; 4.2978E+00 4.9222E+02 6.3988E+02 9.5982E+02;...
4.3013E+00 4.6018E+02 5.9824E+02 8.9736E+02; 4.3067E+00 4.3250E+02 5.6225E+02 8.4337E+02; 4.3124E+00 4.0898E+02 5.3168E+02 7.9752E+02;...
4.3221E+00 3.7984E+02 4.9379E+02 7.4068E+02; 4.3351E+00 3.8592E+02 5.0170E+02 7.5255E+02; 4.3480E+00 3.9839E+02 5.1791E+02 7.7687E+02;...
4.3596E+00 3.7773E+02 4.9105E+02 7.3657E+02; 4.3692E+00 3.6013E+02 4.6817E+02 7.0226E+02; 4.3834E+00 3.8410E+02 4.9933E+02 7.4900E+02;...
4.3926E+00 4.1092E+02 5.3419E+02 8.0129E+02; 4.4018E+00 4.0240E+02 5.2312E+02 7.8468E+02; 4.4088E+00 3.3447E+02 4.3481E+02 6.5221E+02;...
4.4171E+00 2.5545E+02 3.3209E+02 4.9813E+02; 4.4238E+00 2.1344E+02 2.7747E+02 4.1621E+02; 4.4305E+00 1.7919E+02 2.3295E+02 3.4942E+02;...
4.4421E+00 1.4275E+02 1.8557E+02 2.7836E+02; 4.4604E+00 1.1618E+02 1.5104E+02 2.2656E+02; 4.4796E+00 1.0112E+02 1.3145E+02 1.9718E+02;...
4.4971E+00 9.6166E+01 1.2502E+02 1.8752E+02; 4.5179E+00 9.3516E+01 1.2157E+02 1.8236E+02; 4.5297E+00 9.2997E+01 1.2090E+02 1.8135E+02;...
4.5381E+00 9.2997E+01 1.2090E+02 1.8135E+02; 4.5520E+00 9.3516E+01 1.2157E+02 1.8236E+02; 4.5651E+00 9.7166E+01 1.2632E+02 1.8947E+02;...
```

```
    4.5774E+00 1.0221E+02 1.3288E+02 1.9932E+02; 4.5842E+00 1.0564E+02 1.3733E+02 2.0599E+02; 4.5938E+00 1.0930E+02 1.4209E+02 2.1314E+02;...
    4.6058E+00 1.1052E+02 1.4367E+02 2.1551E+02; 4.6156E+00 1.0927E+02 1.4204E+02 2.1307E+02; 4.6227E+00 1.0510E+02 1.3663E+02 2.0495E+02;...
    4.6295E+00 1.0161E+02 1.3210E+02 1.9814E+02; 4.6367E+00 9.8238E+01 1.2771E+02 1.9156E+02; 4.6481E+00 9.3920E+01 1.2210E+02 1.8314E+02;...
    4.6589E+00 9.2884E+01 1.2075E+02 1.8112E+02; 4.6688E+00 9.7666E+01 1.2697E+02 1.9045E+02; 4.6754E+00 1.0273E+02 1.3355E+02 2.0033E+02;...
    4.6844E+00 1.1008E+02 1.4310E+02 2.1466E+02; 4.6924E+00 1.1867E+02 1.5427E+02 2.3140E+02; 4.7014E+00 1.2704E+02 1.6515E+02 2.4773E+02;...
    4.7074E+00 1.3609E+02 1.7691E+02 2.6537E+02; 4.7178E+00 1.4990E+02 1.9487E+02 2.9230E+02; 4.7270E+00 1.6373E+02 2.1285E+02 3.1927E+02;...
    4.7397E+00 1.8537E+02 2.4097E+02 3.6146E+02; 4.7486E+00 2.1561E+02 2.8029E+02 4.2044E+02; 4.7576E+00 2.5227E+02 3.2795E+02 4.9193E+02;...
    4.7684E+00 2.9868E+02 3.8828E+02 5.8242E+02; 4.7747E+00 3.4407E+02 4.4729E+02 6.7093E+02; 4.7812E+00 3.8903E+02 5.0573E+02 7.5860E+02;...
    4.7875E+00 4.5130E+02 5.8668E+02 8.8003E+02; 4.7952E+00 5.4504E+02 7.0855E+02 1.0000E+03; 4.8010E+00 6.0767E+02 7.8997E+02 1.0000E+03;...
    4.8066E+00 6.8462E+02 8.9000E+02 1.0000E+03; 4.8092E+00 7.2329E+02 9.4028E+02 1.0000E+03; 4.8120E+00 7.5708E+02 9.8421E+02 1.0000E+03;...
    4.8161E+00 8.1538E+02 1.0000E+03 1.0000E+03; 4.8226E+00 9.2308E+02 1.0000E+03 1.0000E+03; 4.8300E+00 1.0000E+03 1.0000E+03 1.0000E+03;...
    4.8400E+00 1.0000E+03 1.0000E+03 1.0000E+03; 4.8587E+00 1.0000E+03 1.0000E+03 1.0000E+03; 4.9097E+00 1.0000E+03 1.0000E+03 1.0000E+03;...
    4.9400E+00 1.0000E+03 1.0000E+03 1.0000E+03; 4.9586E+00 1.0000E+03 1.0000E+03 1.0000E+03; 4.9756E+00 8.4615E+02 1.0000E+03 1.0000E+03;...
    4.9832E+00 7.7692E+02 1.0000E+03 1.0000E+03; 4.9875E+00 7.4902E+02 9.7372E+02 1.0000E+03; 4.9935E+00 7.1784E+02 9.3319E+02 1.0000E+03;...
    5.0023E+00 6.7369E+02 8.7583E+02 1.0000E+03; 5.0212E+00 6.3493E+02 8.2553E+02 1.0000E+03; 5.0391E+00 6.4659E+02 8.4088E+02 1.0000E+03;...
    5.0545E+00 6.5791E+02 8.5583E+02 1.0000E+03; 5.0638E+00 6.1900E+02 8.0540E+02 1.0000E+03; 5.0680E+00 5.5329E+02 7.2002E+02 1.0000E+03;...
    5.0739E+00 4.7401E+02 6.1696E+02 9.2640E+02; 5.0800E+00 3.7938E+02 4.9389E+02 7.4175E+02; 5.0866E+00 3.1395E+02 4.0881E+02 6.1411E+02;...
    5.0952E+00 2.5389E+02 3.3072E+02 4.9695E+02; 5.1084E+00 2.0633E+02 2.6891E+02 4.0425E+02; 5.1210E+00 1.8641E+02 2.4309E+02 3.6563E+02;...
    5.1351E+00 1.8826E+02 2.4568E+02 3.6974E+02; 5.1482E+00 2.2679E+02 2.9615E+02 4.4595E+02; 5.1550E+00 2.6043E+02 3.4023E+02 5.1252E+02;...
    5.1595E+00 2.9156E+02 3.8102E+02 5.7412E+02; 5.1642E+00 3.2034E+02 4.1875E+02 6.3115E+02; 5.1682E+00 3.4519E+02 4.5139E+02 6.8055E+02;...
    5.1781E+00 3.5957E+02 4.7045E+02 7.0961E+02; 5.1847E+00 3.4603E+02 4.5298E+02 6.8357E+02; 5.1954E+00 3.2944E+02 4.3156E+02 6.5164E+02;...
    5.2055E+00 3.3366E+02 4.3739E+02 6.6085E+02; 5.2138E+00 3.4914E+02 4.5798E+02 6.9234E+02; 5.2208E+00 3.6640E+02 4.8091E+02 7.2738E+02;...
    5.2296E+00 3.9177E+02 5.1453E+02 7.7865E+02; 5.2354E+00 4.0550E+02 5.3286E+02 8.0679E+02; 5.2433E+00 4.0756E+02 5.3590E+02 8.1185E+02;...
    5.2517E+00 3.9166E+02 5.1534E+02 7.8116E+02; 5.2587E+00 3.5831E+02 4.7180E+02 7.1556E+02; 5.2683E+00 3.3155E+02 4.3692E+02 6.6315E+02;...
    5.2789E+00 3.2360E+02 4.2684E+02 6.4839E+02; 5.2903E+00 3.3091E+02 4.3692E+02 6.6427E+02; 5.3008E+00 3.4323E+02 4.5364E+02 6.9029E+02;...
    5.3092E+00 3.5956E+02 4.7569E+02 7.2447E+02; 5.3234E+00 3.8106E+02 5.0478E+02 7.6962E+02; 5.3377E+00 3.9959E+02 5.3005E+02 8.0911E+02;...
    5.3513E+00 3.9875E+02 5.2966E+02 8.0948E+02; 5.3630E+00 3.7824E+02 5.0308E+02 7.6973E+02; 5.3769E+00 3.5212E+02 4.6900E+02 7.1847E+02;...
    5.3905E+00 3.4121E+02 4.5513E+02 6.9810E+02; 5.4047E+00 3.5998E+02 4.8086E+02 7.3851E+02; 5.4164E+00 3.9804E+02 5.3242E+02 8.1865E+02;...
    5.4294E+00 4.4348E+02 5.9404E+02 9.1453E+02; 5.4420E+00 4.6787E+02 6.2764E+02 9.6748E+02; 5.4561E+00 5.0167E+02 6.7410E+02 1.0000E+03;...
    5.4776E+00 5.7959E+02 7.8040E+02 1.0000E+03; 5.4947E+00 6.1694E+02 8.3219E+02 1.0000E+03; 5.5052E+00 5.7136E+02 7.7172E+02 1.0000E+03;...
    5.5136E+00 4.9816E+02 6.7363E+02 1.0000E+03; 5.5271E+00 4.4811E+02 6.0678E+02 9.4299E+02; 5.5410E+00 4.6541E+02 6.3111E+02 9.8203E+02;...
    5.5519E+00 4.9859E+02 6.7697E+02 1.0000E+03; 5.5651E+00 5.1011E+02 6.9348E+02 1.0000E+03; 5.5729E+00 4.7243E+02 6.4285E+02 1.0000E+03;...
    5.5782E+00 4.2846E+02 5.8341E+02 9.1107E+02; 5.5839E+00 3.7967E+02 5.1732E+02 8.0833E+02; 5.5912E+00 3.3024E+02 4.5032E+02 7.0413E+02;...
    5.5993E+00 2.9970E+02 4.0905E+02 6.4012E+02; 5.6121E+00 2.8073E+02 3.8363E+02 6.0097E+02; 5.6242E+00 2.8804E+02 3.9412E+02 6.1807E+02;...
    5.6379E+00 3.1806E+02 4.3571E+02 6.8403E+02; 5.6452E+00 3.5199E+02 4.8261E+02 7.5821E+02; 5.6516E+00 3.9121E+02 5.3673E+02 8.4372E+02;...
    5.6562E+00 4.2596E+02 5.8470E+02 9.1953E+02; 5.6607E+00 4.6848E+02 6.4338E+02 1.0000E+03; 5.6671E+00 5.3968E+02 7.4157E+02 1.0000E+03;...
    5.6720E+00 6.0503E+02 8.3175E+02 1.0000E+03; 5.6763E+00 6.8668E+02 9.4437E+02 1.0000E+03; 5.6784E+00 7.1880E+02 9.8883E+02 1.0000E+03;...
    5.6814E+00 7.7750E+02 1.0000E+03 1.0000E+03; 5.6885E+00 9.0768E+02 1.0000E+03 1.0000E+03; 5.7000E+00 1.0000E+03 1.0000E+03 1.0000E+03;...
    5.7126E+00 1.0000E+03 1.0000E+03 1.0000E+03; 5.7502E+00 1.0000E+03 1.0000E+03 1.0000E+03; 5.7700E+00 1.0000E+03 1.0000E+03 1.0000E+03;...
    5.7818E+00 9.3692E+02 1.0000E+03 1.0000E+03; 5.7870E+00 8.5000E+02 1.0000E+03 1.0000E+03; 5.7911E+00 7.7051E+02 1.0000E+03 1.0000E+03;...
    5.7940E+00 7.1994E+02 1.0000E+03 1.0000E+03; 5.7961E+00 6.8831E+02 9.5629E+02 1.0000E+03; 5.8005E+00 6.0464E+02 8.4040E+02 1.0000E+03;...
    5.8136E+00 4.4000E+02 6.1209E+02 9.7390E+02; 5.8350E+00 2.8686E+02 3.9954E+02 6.3640E+02; 5.8621E+00 2.2484E+02 3.1360E+02 5.0013E+02;...
    5.8921E+00 2.0016E+02 2.7956E+02 4.4637E+02; 5.9327E+00 2.0486E+02 2.8647E+02 4.5789E+02; 5.9656E+00 2.2016E+02 3.0810E+02 4.9278E+02;...
    6.0280E+00 2.5887E+02 3.6238E+02 5.7977E+02; 6.0769E+00 3.1031E+02 4.3443E+02 6.9508E+02; 6.1395E+00 4.1120E+02 5.7568E+02 9.2108E+02;...
    6.1600E+00 4.5786E+02 6.4100E+02 1.0000E+03; 6.1823E+00 5.2017E+02 7.2824E+02 1.0000E+03; 6.2000E+00 5.8714E+02 8.2200E+02 1.0000E+03;...
    6.2125E+00 6.4071E+02 8.9700E+02 1.0000E+03; 6.2214E+00 6.8356E+02 9.5699E+02 1.0000E+03; 6.2267E+00 7.1071E+02 9.9500E+02 1.0000E+03;...
    6.2334E+00 7.5000E+02 1.0000E+03 1.0000E+03; 6.2498E+00 8.4286E+02 1.0000E+03 1.0000E+03; 6.2650E+00 9.5000E+02 1.0000E+03 1.0000E+03;...
    6.2800E+00 1.0000E+03 1.0000E+03 1.0000E+03; 6.3103E+00 1.0000E+03 1.0000E+03 1.0000E+03; 6.3500E+00 1.0000E+03 1.0000E+03 1.0000E+03;...
    ];
end %end of create global TableOfBBOabsorptionCoef_iwfjp

%absorption coefficient calculation
AbsCoef_min=interp1(TableOfBBOabsorptionCoef_iwfjp(:,1), TableOfBBOabsorptionCoef_iwfjp(:,2), inputValues, 'linear', 'extrap');
AbsCoef_typ=interp1(TableOfBBOabsorptionCoef_iwfjp(:,1), TableOfBBOabsorptionCoef_iwfjp(:,3), inputValues, 'linear', 'extrap');
AbsCoef_max=interp1(TableOfBBOabsorptionCoef_iwfjp(:,1), TableOfBBOabsorptionCoef_iwfjp(:,4), inputValues, 'linear', 'extrap');
% end of absorption coefficient calculation

%refraction index calculation
RefracInd_o=sqrt(1+ inputValues.^2 *0.90291 ./ (inputValues.^2 -0.003926) +...
 inputValues.^2 *0.83155 ./ (inputValues.^2 -0.018786) +...
 inputValues.^2 *0.76536 ./ (inputValues.^2 -60.01));

RefracInd_e=sqrt(1+ inputValues.^2 *1.151075 ./ (inputValues.^2 -0.007142) +...
 inputValues.^2 *0.21803 ./ (inputValues.^2 -0.02259) +...
 inputValues.^2 *0.656 ./ (inputValues.^2 -263));
% end of refraction index calculation

% data to return
ValueToReturn=[inputValues, AbsCoef_min, AbsCoef_typ, AbsCoef_max, RefracInd_o, RefracInd_e];

end % end function
```

An example for use:

```
SomeData=AbsorptionBBO_v0(0.30:0.002:2.0); % getting data for 0.3-2 mkm
figure(287);clf();
subplot(3,1,1);
 semilogy(SomeData(:,1),SomeData(:,2:4)); grid on;
 xlabel('Wavelength, mkm');
 ylabel('Attenuation, -dB/cm');
subplot(3,1,2);
InPower=2; % input power W (no reflection losses)
CrystalLength=5.5;% crystal length mm
 semilogy(SomeData(:,1), InPower.*(1-10.^(-SomeData(:,2:4)/10.* CrystalLength/10 )) );
 title('Absorbed power: '+string(InPower)+'W input, length '+string(CrystalLength)+'mm');
 xlabel('Wavelength, mkm'); ylabel('Absorbed power, W'); grid on;
subplot(3,1,3);
 plot(SomeData(:,1),SomeData(:,5:6));
 ylabel('Refractive index'); xlabel('Wavelength, mkm'); grid on;
```

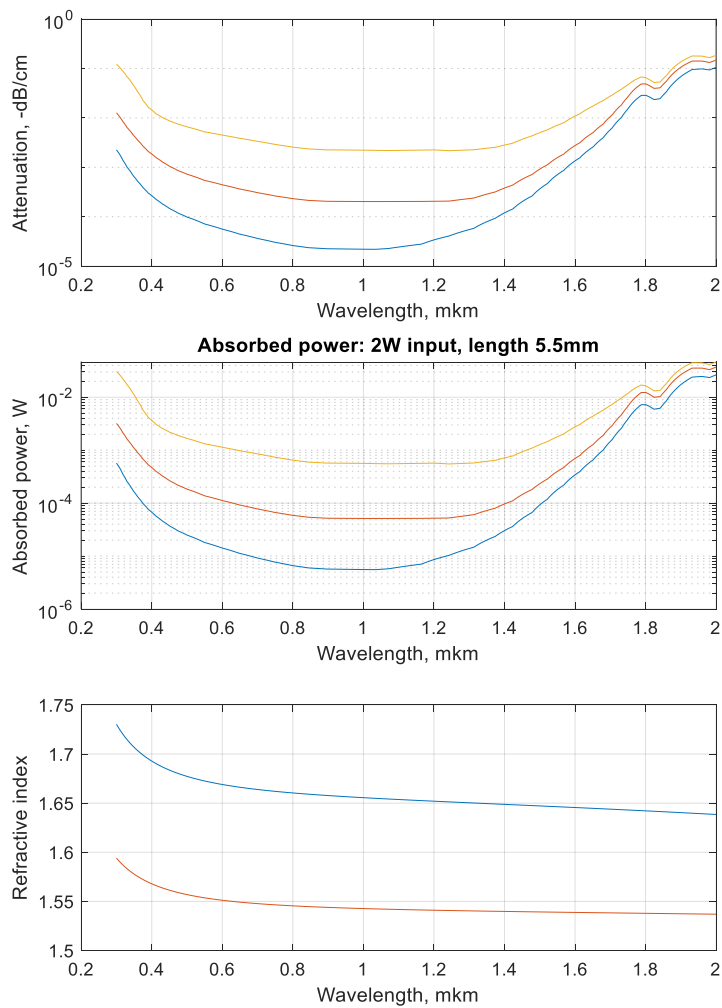

# Section 3.

SCILAB function for calculating absorption coefficient and refractive indices. Function was written and tested using SCILAB 6.1.1. Users can make their own modifications.

```
function ValueToReturn=AbsorptionBBO_v0(Wavelength_mkm)
// Function returns minimal typical and maximal values of attenuation
//    as power loss 'dB/cm', and refractive indices of o and e polarizations
// T=10^(-dB*L/10) - transmittance
//
// INPUT is a vector wavelength in mkm
// OUTPUT=[wavelength; min; typ; max; o; e]
// Attenuation values are clipped at 150 dB/cm in UV, 1000 dB/cm in IR
//
// Attenuation coefficient is valid 0.188-6.22 mkm
// Sellmeier equation valid 0.188-5.200 mkm, Opt.Mater.Express,8,1410-1418(2018)
//
// date: 2021.11.01
//
// Example:
// SomeData=AbsorptionBBO_v0([0.532; 1.064; 1.550]);

//checking input is valid
if exists('Wavelength_mkm', 'l')==0 || isempty(Wavelength_mkm) || type(Wavelength_mkm)~=1
    mprintf('%s\n', 'Warning, input is empty or not a number in function ''AbsorptionBBO_v0''');
    ValueToReturn=[];
    return
end

inputSize=size(Wavelength_mkm);
if inputSize(2)<inputSize(1) && min(inputSize)==1
    inputValues=Wavelength_mkm;
elseif min(inputSize)==1
    inputValues=Wavelength_mkm.';
else
    mprintf('%s\n', 'Warning, input must be a vector in function ''AbsorptionBBO_v0''');
    ValueToReturn=[];
    return
end // end of input checking

//creates variable if not yet exists; otherwise uses existing global
if ~isglobal('TableOfBBOabsorptionCoef_iwfjp')
    global TableOfBBOabsorptionCoef_iwfjp
if isempty(TableOfBBOabsorptionCoef_iwfjp)
    mprintf('%s\n', 'global variable is created in function ''AbsorptionBBO_v0''');
TableOfBBOabsorptionCoef_iwfjp=[...
1.8500E-01 1.5000E+02 1.5000E+02 1.5000E+02; 1.8530E-01 1.5000E+02 1.5000E+02 1.5000E+02; 1.8590E-01 1.5000E+02 1.5000E+02 1.5000E+02;...
1.8640E-01 1.3157E+02 1.5000E+02 1.5000E+02; 1.8679E-01 1.1200E+02 1.5000E+02 1.5000E+02; 1.8701E-01 1.0324E+02 1.5000E+02 1.5000E+02;...
1.8745E-01 8.6507E+01 1.5000E+02 1.5000E+02; 1.8810E-01 6.7513E+01 1.2760E+02 1.5000E+02; 1.8870E-01 5.4510E+01 1.0329E+02 1.5000E+02;...
1.8930E-01 4.3283E+01 8.2200E+01 1.5000E+02; 1.9021E-01 3.0908E+01 5.8907E+01 1.5000E+02; 1.9100E-01 2.2154E+01 4.2600E+01 1.2808E+02;...
1.9150E-01 1.7920E+01 3.5000E+01 1.0553E+02; 1.9204E-01 1.3984E+01 2.8198E+01 8.5532E+01; 1.9399E-01 6.2526E+00 1.3500E+01 4.1499E+01;...
1.9597E-01 2.8122E+00 6.5564E+00 2.0477E+01; 1.9746E-01 1.5182E+00 3.7480E+00 1.1854E+01; 1.9913E-01 8.0279E-01 2.0557E+00 6.5599E+00;...
2.0059E-01 4.7717E-01 1.2426E+00 3.9932E+00; 2.0241E-01 2.8270E-01 7.4901E-01 2.4441E+00; 2.0459E-01 1.7431E-01 4.7702E-01 1.6080E+00;...
2.0748E-01 1.0629E-01 3.0757E-01 1.0975E+00; 2.1239E-01 6.9493E-02 2.1715E-01 8.4060E-01; 2.1978E-01 4.6592E-02 1.5449E-01 6.4696E-01;...
2.2664E-01 3.3634E-02 1.1258E-01 5.0944E-01; 2.3555E-01 2.3632E-02 7.8577E-02 4.0108E-01; 2.4777E-01 1.4629E-02 4.9864E-02 2.9194E-01;...
2.6379E-01 7.9651E-03 3.1086E-02 2.1196E-01; 2.8009E-01 4.3116E-03 1.9992E-02 1.6043E-01; 2.9662E-01 2.4107E-03 1.3427E-02 1.2846E-01;...
3.1798E-01 1.3960E-03 8.4544E-03 8.8707E-02; 3.3000E-01 1.0057E-03 6.3708E-03 6.8134E-02; 3.4702E-01 6.8104E-04 4.5391E-03 4.6718E-02;...
3.6000E-01 5.2113E-04 3.5478E-03 3.3776E-02; 3.7556E-01 3.8512E-04 2.6533E-03 2.2261E-02; 3.9000E-01 3.0126E-04 2.0939E-03 1.6431E-02;...
4.1156E-01 2.2589E-04 1.5837E-03 1.2194E-02; 4.3000E-01 1.8176E-04 1.2859E-03 1.0158E-02; 4.5088E-01 1.4566E-04 1.0425E-03 8.6360E-03;...
4.7500E-01 1.1791E-04 8.5507E-04 7.4511E-03; 4.9996E-01 9.8396E-05 7.2497E-04 6.5879E-03; 5.4934E-01 7.1447E-05 5.4365E-04 5.2204E-03;...
5.9999E-01 5.5989E-05 4.4213E-04 4.4877E-03; 6.4824E-01 4.4599E-05 3.6501E-04 3.8705E-03; 7.0098E-01 3.6078E-05 3.0532E-04 3.3459E-03;...
7.5036E-01 3.0249E-05 2.6250E-04 2.8987E-03; 7.9973E-01 2.6049E-05 2.3197E-04 2.5586E-03; 8.4462E-01 2.3588E-05 2.1362E-04 2.3516E-03;...
8.9624E-01 2.2470E-05 2.0499E-04 2.2551E-03; 9.5571E-01 2.2184E-05 2.0358E-04 2.2378E-03; 1.0017E+00 2.1964E-05 2.0280E-04 2.2221E-03;...
1.0350E+00 2.1884E-05 2.0280E-04 2.2007E-03; 1.0714E+00 2.2734E-05 2.0290E-04 2.1844E-03; 1.1651E+00 2.7907E-05 2.0384E-04 2.2298E-03;...
1.2000E+00 3.3916E-05 2.0511E-04 2.2492E-03; 1.2426E+00 4.0699E-05 2.0689E-04 2.1689E-03; 1.3130E+00 5.8122E-05 2.3855E-04 2.2640E-03;...
1.3735E+00 9.2663E-05 3.1658E-04 2.5361E-03; 1.4233E+00 1.4307E-04 4.3615E-04 3.0670E-03; 1.4786E+00 2.6365E-04 7.1681E-04 4.3400E-03;...
1.5218E+00 4.6930E-04 1.1256E-03 5.7321E-03; 1.5681E+00 8.6597E-04 1.8544E-03 7.9829E-03; 1.6139E+00 1.5595E-03 3.1450E-03 1.2130E-02;...
1.6517E+00 2.5993E-03 5.0696E-03 1.7459E-02; 1.6798E+00 3.9662E-03 7.5127E-03 2.2440E-02; 1.7056E+00 6.2740E-03 1.1536E-02 2.8993E-02;...
1.7318E+00 1.0280E-02 1.8379E-02 3.8158E-02; 1.7493E+00 1.5030E-02 2.6322E-02 4.6357E-02; 1.7687E+00 2.2052E-02 3.8021E-02 5.8097E-02;...
1.7878E+00 2.8513E-02 4.8691E-02 6.7548E-02; 1.8008E+00 2.8617E-02 4.8666E-02 6.4932E-02; 1.8100E+00 2.6763E-02 4.5368E-02 5.9928E-02;...
1.8237E+00 2.3482E-02 3.9545E-02 5.2060E-02; 1.8413E+00 2.4424E-02 4.0614E-02 5.3259E-02; 1.8590E+00 3.3688E-02 5.4954E-02 7.1658E-02;...
1.8825E+00 5.3320E-02 8.4623E-02 1.0947E-01; 1.9048E+00 7.2716E-02 1.1177E-01 1.4329E-01; 1.9320E+00 9.5222E-02 1.4131E-01 1.7940E-01;...
1.9602E+00 9.8210E-02 1.4138E-01 1.7803E-01; 1.9814E+00 9.3455E-02 1.3210E-01 1.6554E-01; 2.0096E+00 1.1107E-01 1.5570E-01 1.9474E-01;...
2.0308E+00 1.3972E-01 1.9510E-01 2.4390E-01; 2.0484E+00 1.6900E-01 2.3510E-01 2.9388E-01; 2.0672E+00 1.9493E-01 2.6980E-01 3.3725E-01;...
2.0860E+00 2.1601E-01 2.9700E-01 3.7125E-01; 2.1084E+00 2.1792E-01 2.9700E-01 3.7125E-01; 2.1331E+00 2.5609E-01 3.4540E-01 4.3175E-01;...
2.1530E+00 3.2676E-01 4.3630E-01 5.4538E-01; 2.1695E+00 4.5231E-01 5.9840E-01 7.4800E-01; 2.1848E+00 6.2573E-01 8.2070E-01 1.0259E+00;...
```

```
2.2001E+00 8.4183E-01 1.0952E+00 1.3690E+00; 2.2105E+00 1.0617E+00 1.3721E+00 1.7151E+00; 2.2210E+00 1.2955E+00 1.6641E+00 2.0801E+00;...
2.2304E+00 1.5187E+00 1.9397E+00 2.4246E+00; 2.2410E+00 1.7266E+00 2.1924E+00 2.7405E+00; 2.2516E+00 1.8633E+00 2.3523E+00 2.9404E+00;...
2.2619E+00 2.0679E+00 2.5966E+00 3.2458E+00; 2.2691E+00 2.2740E+00 2.8427E+00 3.5534E+00; 2.2783E+00 2.5010E+00 3.1120E+00 3.8900E+00;...
2.2868E+00 2.8267E+00 3.5009E+00 4.3761E+00; 2.2996E+00 2.8873E+00 3.5573E+00 4.4466E+00; 2.3093E+00 2.9699E+00 3.6425E+00 4.5531E+00;...
2.3187E+00 3.1374E+00 3.8326E+00 4.7908E+00; 2.3284E+00 3.3803E+00 4.1146E+00 5.1433E+00; 2.3367E+00 3.6252E+00 4.3991E+00 5.4989E+00;...
2.3466E+00 3.8118E+00 4.6124E+00 5.7655E+00; 2.3553E+00 3.7316E+00 4.5045E+00 5.6306E+00; 2.3637E+00 3.5810E+00 4.3141E+00 5.3926E+00;...
2.3745E+00 3.6299E+00 4.3654E+00 5.4568E+00; 2.3845E+00 3.7635E+00 4.5205E+00 5.6506E+00; 2.3972E+00 3.7801E+00 4.5374E+00 5.6718E+00;...
2.4031E+00 3.2776E+00 3.9339E+00 4.9174E+00; 2.4120E+00 2.3857E+00 2.8646E+00 3.5808E+00; 2.4192E+00 1.8298E+00 2.1992E+00 2.7490E+00;...
2.4348E+00 1.5497E+00 1.8663E+00 2.3329E+00; 2.4480E+00 1.6322E+00 1.9707E+00 2.4634E+00; 2.4596E+00 1.7867E+00 2.1632E+00 2.7040E+00;...
2.4700E+00 1.8998E+00 2.3068E+00 2.8835E+00; 2.4822E+00 2.0490E+00 2.4957E+00 3.1196E+00; 2.4889E+00 2.1429E+00 2.6167E+00 3.2709E+00;...
2.4986E+00 2.2402E+00 2.7435E+00 3.4294E+00; 2.5100E+00 2.4246E+00 2.9787E+00 3.7234E+00; 2.5193E+00 2.5152E+00 3.0992E+00 3.8740E+00;...
2.5300E+00 2.6608E+00 3.2881E+00 4.1101E+00; 2.5402E+00 2.8254E+00 3.5009E+00 4.3761E+00; 2.5494E+00 2.9454E+00 3.6581E+00 4.5726E+00;...
2.5597E+00 3.1330E+00 3.8993E+00 4.8741E+00; 2.5712E+00 3.3915E+00 4.2287E+00 5.2859E+00; 2.5809E+00 3.6344E+00 4.5374E+00 5.6717E+00;...
2.5886E+00 3.8160E+00 4.7676E+00 5.9595E+00; 2.5988E+00 4.1426E+00 5.1775E+00 6.4717E+00; 2.6100E+00 4.3860E+00 5.4816E+00 6.8511E+00;...
2.6202E+00 4.5063E+00 5.6308E+00 7.0359E+00; 2.6328E+00 4.7423E+00 5.9235E+00 7.3989E+00; 2.6395E+00 4.9080E+00 6.1281E+00 7.6516E+00;...
2.6462E+00 5.1996E+00 6.4893E+00 8.0989E+00; 2.6563E+00 5.6023E+00 6.9877E+00 8.7156E+00; 2.6649E+00 5.9186E+00 7.3772E+00 9.1952E+00;...
2.6738E+00 5.9836E+00 7.4523E+00 9.2816E+00; 2.6824E+00 5.6528E+00 7.0339E+00 8.7524E+00; 2.6956E+00 4.7037E+00 5.8460E+00 7.2657E+00;...
2.7056E+00 4.3423E+00 5.3906E+00 6.6920E+00; 2.7141E+00 4.2169E+00 5.2292E+00 6.4845E+00; 2.7229E+00 4.2650E+00 5.2825E+00 6.5427E+00;...
2.7354E+00 4.5854E+00 5.6709E+00 7.0133E+00; 2.7464E+00 4.9273E+00 6.0848E+00 7.5142E+00; 2.7553E+00 5.2088E+00 6.4239E+00 7.9224E+00;...
2.7636E+00 5.4957E+00 6.7689E+00 8.3370E+00; 2.7733E+00 5.9015E+00 7.2582E+00 8.9267E+00; 2.7843E+00 6.2757E+00 7.7058E+00 9.4618E+00;...
2.7959E+00 6.6054E+00 8.0966E+00 9.9244E+00; 2.8077E+00 6.9188E+00 8.4656E+00 1.0358E+01; 2.8197E+00 7.3600E+00 8.9894E+00 1.0980E+01;...
2.8301E+00 7.4042E+00 9.0281E+00 1.1008E+01; 2.8432E+00 7.2240E+00 8.7930E+00 1.0703E+01; 2.8530E+00 7.2433E+00 8.8032E+00 1.0699E+01;...
2.8621E+00 7.3360E+00 8.9037E+00 1.0806E+01; 2.8728E+00 7.6809E+00 9.3095E+00 1.1283E+01; 2.8824E+00 8.2230E+00 9.9542E+00 1.2050E+01;...
2.8901E+00 8.7495E+00 1.0580E+01 1.2794E+01; 2.8992E+00 9.1253E+00 1.1022E+01 1.3313E+01; 2.9101E+00 9.6453E+00 1.1637E+01 1.4039E+01;...
2.9212E+00 1.0331E+01 1.2450E+01 1.5004E+01; 2.9329E+00 1.0774E+01 1.2970E+01 1.5615E+01; 2.9421E+00 1.0783E+01 1.2970E+01 1.5602E+01;...
2.9520E+00 1.0507E+01 1.2630E+01 1.5182E+01; 2.9602E+00 9.8323E+00 1.1813E+01 1.4191E+01; 2.9725E+00 9.9276E+00 1.1921E+01 1.4315E+01;...
2.9795E+00 1.0446E+01 1.2540E+01 1.5053E+01; 2.9907E+00 1.1112E+01 1.3336E+01 1.6005E+01; 2.9998E+00 1.1113E+01 1.3336E+01 1.6004E+01;...
3.0093E+00 1.0822E+01 1.2987E+01 1.5586E+01; 3.0184E+00 1.1730E+01 1.4077E+01 1.6898E+01; 3.0273E+00 1.2639E+01 1.5170E+01 1.8215E+01;...
3.0369E+00 1.3715E+01 1.6464E+01 1.9778E+01; 3.0484E+00 1.4776E+01 1.7742E+01 2.1328E+01; 3.0598E+00 1.5522E+01 1.8643E+01 2.2430E+01;...
3.0709E+00 1.6242E+01 1.9514E+01 2.3500E+01; 3.0801E+00 1.5682E+01 1.8847E+01 2.2721E+01; 3.0895E+00 1.5373E+01 1.8483E+01 2.2307E+01;...
3.1000E+00 1.6009E+01 1.9256E+01 2.3267E+01; 3.1080E+00 1.5942E+01 1.9182E+01 2.3204E+01; 3.1139E+00 1.4885E+01 1.7916E+01 2.1697E+01;...
3.1237E+00 1.3412E+01 1.6151E+01 1.9587E+01; 3.1327E+00 1.2759E+01 1.5373E+01 1.8673E+01; 3.1445E+00 1.3656E+01 1.6464E+01 2.0037E+01;...
3.1550E+00 1.4884E+01 1.7958E+01 2.1900E+01; 3.1670E+00 1.5624E+01 1.8865E+01 2.3060E+01; 3.1790E+00 1.6482E+01 1.9917E+01 2.4403E+01;...
3.1878E+00 1.7514E+01 2.1179E+01 2.6003E+01; 3.1964E+00 1.9249E+01 2.3294E+01 2.8660E+01; 3.2068E+00 2.1564E+01 2.6117E+01 3.2211E+01;...
3.2176E+00 2.4681E+01 2.9919E+01 3.7002E+01; 3.2316E+00 2.9401E+01 3.5680E+01 4.4268E+01; 3.2435E+00 3.5235E+01 4.2805E+01 5.3272E+01;...
3.2540E+00 4.1919E+01 5.0973E+01 6.3613E+01; 3.2615E+00 4.8339E+01 5.8827E+01 7.3590E+01; 3.2704E+00 5.6486E+01 6.8802E+01 8.6282E+01;...
3.2801E+00 6.4664E+01 7.8836E+01 9.9135E+01; 3.2896E+00 7.6027E+01 9.2780E+01 1.1700E+02; 3.3007E+00 8.7526E+01 1.0693E+02 1.3525E+02;...
3.3098E+00 9.5709E+01 1.1704E+02 1.4847E+02; 3.3196E+00 1.0303E+02 1.2612E+02 1.6045E+02; 3.3279E+00 1.1013E+02 1.3493E+02 1.7215E+02;...
3.3377E+00 1.2529E+02 1.5367E+02 1.9663E+02; 3.3467E+00 1.4354E+02 1.7625E+02 2.2624E+02; 3.3595E+00 1.6512E+02 2.0302E+02 2.6158E+02;...
3.3720E+00 1.8921E+02 2.3295E+02 3.0130E+02; 3.3821E+00 2.1581E+02 2.6602E+02 3.4532E+02; 3.3936E+00 2.3823E+02 2.9405E+02 3.8313E+02;...
3.4057E+00 2.6325E+02 3.2537E+02 4.2555E+02; 3.4166E+00 2.8597E+02 3.5389E+02 4.6450E+02; 3.4253E+00 2.9751E+02 3.6857E+02 4.8527E+02;...
3.4346E+00 3.0556E+02 3.7894E+02 5.0036E+02; 3.4410E+00 3.1018E+02 3.8501E+02 5.0964E+02; 3.4488E+00 3.2169E+02 3.9967E+02 5.3042E+02;...
3.4580E+00 3.3522E+02 4.1687E+02 5.5472E+02; 3.4630E+00 3.5696E+02 4.4424E+02 5.9241E+02; 3.4697E+00 3.8588E+02 4.8062E+02 6.4236E+02;...
3.4775E+00 4.0438E+02 5.0410E+02 6.7539E+02; 3.4840E+00 4.1511E+02 5.1791E+02 6.9555E+02; 3.4919E+00 3.9540E+02 4.9379E+02 6.6489E+02;...
3.5010E+00 3.4233E+02 4.2795E+02 5.7789E+02; 3.5089E+00 3.0595E+02 3.8287E+02 5.1853E+02; 3.5182E+00 2.8087E+02 3.5193E+02 4.7826E+02;...
3.5341E+00 2.7013E+02 3.3902E+02 4.6280E+02; 3.5461E+00 2.6841E+02 3.3741E+02 4.6266E+02; 3.5634E+00 2.7657E+02 3.4831E+02 4.8002E+02;...
3.5781E+00 2.8638E+02 3.6129E+02 5.0026E+02; 3.5929E+00 2.8753E+02 3.6330E+02 5.0521E+02; 3.6024E+00 2.8423E+02 3.5957E+02 5.0164E+02;...
3.6106E+00 2.7073E+02 3.4282E+02 4.7955E+02; 3.6187E+00 2.3686E+02 3.0019E+02 4.2094E+02; 3.6255E+00 2.0414E+02 2.5893E+02 3.6385E+02;...
3.6310E+00 1.6431E+02 2.0858E+02 2.9372E+02; 3.6419E+00 1.2666E+02 1.6096E+02 2.2731E+02; 3.6528E+00 1.0681E+02 1.3588E+02 1.9249E+02;...
3.6625E+00 9.6982E+01 1.2352E+02 1.7551E+02; 3.6761E+00 9.4802E+01 1.2090E+02 1.7237E+02; 3.6881E+00 9.8088E+01 1.2523E+02 1.7912E+02;...
3.6970E+00 1.0137E+02 1.2955E+02 1.8579E+02; 3.7068E+00 1.0844E+02 1.3872E+02 1.9944E+02; 3.7164E+00 1.1490E+02 1.4713E+02 2.1206E+02;...
3.7263E+00 1.1599E+02 1.4866E+02 2.1480E+02; 3.7374E+00 1.0935E+02 1.4027E+02 2.0320E+02; 3.7459E+00 1.0309E+02 1.3237E+02 1.9219E+02;...
3.7568E+00 9.9602E+01 1.2801E+02 1.8635E+02; 3.7714E+00 1.0230E+02 1.3163E+02 1.9218E+02; 3.7854E+00 1.0491E+02 1.3513E+02 1.9784E+02;...
3.7955E+00 1.0821E+02 1.3950E+02 2.0468E+02; 3.8041E+00 1.0987E+02 1.4173E+02 2.0835E+02; 3.8155E+00 1.1217E+02 1.4481E+02 2.1327E+02;...
3.8226E+00 1.1509E+02 1.4866E+02 2.1928E+02; 3.8312E+00 1.1742E+02 1.5176E+02 2.2420E+02; 3.8414E+00 1.2057E+02 1.5592E+02 2.3069E+02;...
3.8486E+00 1.2570E+02 1.6263E+02 2.4093E+02; 3.8569E+00 1.2945E+02 1.6757E+02 2.4855E+02; 3.8644E+00 1.3308E+02 1.7235E+02 2.5596E+02;...
3.8760E+00 1.3514E+02 1.7511E+02 2.6044E+02; 3.8876E+00 1.3723E+02 1.7792E+02 2.6498E+02; 3.8980E+00 1.4240E+02 1.8469E+02 2.7538E+02;...
3.9039E+00 1.5002E+02 1.9464E+02 2.9046E+02; 3.9141E+00 1.6328E+02 2.1192E+02 3.1653E+02; 3.9214E+00 1.7660E+02 2.2927E+02 3.4270E+02;...
3.9302E+00 1.9531E+02 2.5363E+02 3.7938E+02; 3.9391E+00 2.1704E+02 2.8192E+02 4.2197E+02; 3.9480E+00 2.4119E+02 3.1336E+02 4.6930E+02;...
3.9585E+00 2.7958E+02 3.6330E+02 5.4437E+02; 3.9674E+00 3.2075E+02 4.1687E+02 6.2489E+02; 3.9764E+00 3.6220E+02 4.7079E+02 7.0592E+02;...
3.9838E+00 4.0033E+02 5.2039E+02 7.8044E+02; 3.9942E+00 4.3585E+02 5.6659E+02 8.4984E+02; 4.0048E+00 4.5932E+02 5.9712E+02 8.9567E+02;...
4.0138E+00 4.8677E+02 6.3280E+02 9.4919E+02; 4.0245E+00 5.1340E+02 6.6742E+02 1.0000E+03; 4.0425E+00 5.1340E+02 6.6742E+02 1.0000E+03;...
4.0624E+00 5.0530E+02 6.5689E+02 9.8533E+02; 4.0730E+00 4.9460E+02 6.4298E+02 9.6447E+02; 4.0854E+00 4.8948E+02 6.3633E+02 9.5449E+02;...
4.0960E+00 4.8677E+02 6.3280E+02 9.4920E+02; 4.1053E+00 4.7681E+02 6.1985E+02 9.2978E+02; 4.1145E+00 4.6409E+02 6.0332E+02 9.0498E+02;...
4.1208E+00 4.4992E+02 5.8490E+02 8.7735E+02; 4.1306E+00 4.4165E+02 5.7415E+02 8.6122E+02; 4.1392E+00 4.5714E+02 5.9428E+02 8.9141E+02;...
4.1472E+00 4.7909E+02 6.2282E+02 9.3422E+02; 4.1533E+00 5.0812E+02 6.6056E+02 9.9083E+02; 4.1596E+00 5.3252E+02 6.9228E+02 1.0000E+03;...
4.1674E+00 5.6166E+02 7.3015E+02 1.0000E+03; 4.1784E+00 5.6749E+02 7.3774E+02 1.0000E+03; 4.1879E+00 5.4106E+02 7.0337E+02 1.0000E+03;...
4.2035E+00 5.2163E+02 6.7811E+02 1.0000E+03; 4.2256E+00 5.4106E+02 7.0337E+02 1.0000E+03; 4.2337E+00 5.5854E+02 7.2610E+02 1.0000E+03;...
4.2433E+00 5.8258E+02 7.5735E+02 1.0000E+03; 4.2495E+00 6.0140E+02 7.8182E+02 1.0000E+03; 4.2575E+00 6.3430E+02 8.2459E+02 1.0000E+03;...
4.2656E+00 6.5116E+02 8.4651E+02 1.0000E+03; 4.2719E+00 6.4089E+02 8.3316E+02 1.0000E+03; 4.2801E+00 6.0813E+02 7.9057E+02 1.0000E+03;...
4.2849E+00 5.7384E+02 7.4599E+02 1.0000E+03; 4.2912E+00 5.3549E+02 6.9614E+02 1.0000E+03; 4.2978E+00 4.9222E+02 6.3988E+02 9.5982E+02;...
4.3013E+00 4.6018E+02 5.9824E+02 8.9736E+02; 4.3067E+00 4.3250E+02 5.6225E+02 8.4337E+02; 4.3124E+00 4.0898E+02 5.3168E+02 7.9752E+02;...
4.3221E+00 3.7984E+02 4.9379E+02 7.4068E+02; 4.3351E+00 3.8592E+02 5.0170E+02 7.5255E+02; 4.3480E+00 3.9839E+02 5.1791E+02 7.7687E+02;...
4.3596E+00 3.7773E+02 4.9105E+02 7.3657E+02; 4.3692E+00 3.6013E+02 4.6817E+02 7.0226E+02; 4.3834E+00 3.8410E+02 4.9933E+02 7.4900E+02;...
4.3926E+00 4.1092E+02 5.3419E+02 8.0129E+02; 4.4018E+00 4.0240E+02 5.2312E+02 7.8468E+02; 4.4088E+00 3.3447E+02 4.3481E+02 6.5221E+02;...
4.4171E+00 2.5545E+02 3.3209E+02 4.9813E+02; 4.4238E+00 2.1344E+02 2.7747E+02 4.1621E+02; 4.4305E+00 1.7919E+02 2.3295E+02 3.4942E+02;...
4.4421E+00 1.4275E+02 1.8557E+02 2.7836E+02; 4.4604E+00 1.1618E+02 1.5104E+02 2.2656E+02; 4.4796E+00 1.0112E+02 1.3145E+02 1.9718E+02;...
4.4971E+00 9.6166E+01 1.2502E+02 1.8752E+02; 4.5179E+00 9.3516E+01 1.2157E+02 1.8236E+02; 4.5297E+00 9.2997E+01 1.2090E+02 1.8135E+02;...
4.5381E+00 9.2997E+01 1.2090E+02 1.8135E+02; 4.5520E+00 9.3516E+01 1.2157E+02 1.8236E+02; 4.5651E+00 9.7166E+01 1.2632E+02 1.8947E+02;...
4.5774E+00 1.0221E+02 1.3288E+02 1.9932E+02; 4.5842E+00 1.0564E+02 1.3733E+02 2.0599E+02; 4.5938E+00 1.0930E+02 1.4209E+02 2.1314E+02;...
4.6058E+00 1.1052E+02 1.4367E+02 2.1551E+02; 4.6156E+00 1.0927E+02 1.4204E+02 2.1307E+02; 4.6227E+00 1.0510E+02 1.3663E+02 2.0495E+02;...
```

```
    4.6295E+00 1.0161E+02 1.3210E+02 1.9814E+02; 4.6367E+00 9.8238E+01 1.2771E+02 1.9156E+02; 4.6481E+00 9.3920E+01 1.2210E+02 1.8314E+02;...
    4.6589E+00 9.2884E+01 1.2075E+02 1.8112E+02; 4.6688E+00 9.7666E+01 1.2697E+02 1.9045E+02; 4.6754E+00 1.0273E+02 1.3355E+02 2.0033E+02;...
    4.6844E+00 1.1008E+02 1.4310E+02 2.1466E+02; 4.6924E+00 1.1867E+02 1.5427E+02 2.3140E+02; 4.7014E+00 1.2704E+02 1.6515E+02 2.4773E+02;...
    4.7074E+00 1.3609E+02 1.7691E+02 2.6537E+02; 4.7178E+00 1.4990E+02 1.9487E+02 2.9230E+02; 4.7270E+00 1.6373E+02 2.1285E+02 3.1927E+02;...
    4.7397E+00 1.8537E+02 2.4097E+02 3.6146E+02; 4.7486E+00 2.1561E+02 2.8029E+02 4.2044E+02; 4.7576E+00 2.5227E+02 3.2795E+02 4.9193E+02;...
    4.7684E+00 2.9868E+02 3.8828E+02 5.8242E+02; 4.7747E+00 3.4407E+02 4.4729E+02 6.7093E+02; 4.7812E+00 3.8903E+02 5.0573E+02 7.5860E+02;...
    4.7875E+00 4.5130E+02 5.8668E+02 8.8003E+02; 4.7952E+00 5.4504E+02 7.0855E+02 1.0000E+03; 4.8010E+00 6.0767E+02 7.8997E+02 1.0000E+03;...
    4.8066E+00 6.8462E+02 8.9000E+02 1.0000E+03; 4.8092E+00 7.2329E+02 9.4028E+02 1.0000E+03; 4.8120E+00 7.5708E+02 9.8421E+02 1.0000E+03;...
    4.8161E+00 8.1538E+02 1.0000E+03 1.0000E+03; 4.8226E+00 9.2308E+02 1.0000E+03 1.0000E+03; 4.8300E+00 1.0000E+03 1.0000E+03 1.0000E+03;...
    4.8400E+00 1.0000E+03 1.0000E+03 1.0000E+03; 4.8587E+00 1.0000E+03 1.0000E+03 1.0000E+03; 4.9097E+00 1.0000E+03 1.0000E+03 1.0000E+03;...
    4.9400E+00 1.0000E+03 1.0000E+03 1.0000E+03; 4.9586E+00 1.0000E+03 1.0000E+03 1.0000E+03; 4.9756E+00 8.4615E+02 1.0000E+03 1.0000E+03;...
    4.9832E+00 7.7692E+02 1.0000E+03 1.0000E+03; 4.9875E+00 7.4902E+02 9.7372E+02 1.0000E+03; 4.9935E+00 7.1784E+02 9.3319E+02 1.0000E+03;...
    5.0023E+00 6.7369E+02 8.7583E+02 1.0000E+03; 5.0212E+00 6.3493E+02 8.2553E+02 1.0000E+03; 5.0391E+00 6.4659E+02 8.4088E+02 1.0000E+03;...
    5.0545E+00 6.5791E+02 8.5583E+02 1.0000E+03; 5.0638E+00 6.1900E+02 8.0540E+02 1.0000E+03; 5.0680E+00 5.5329E+02 7.2002E+02 1.0000E+03;...
    5.0739E+00 4.7401E+02 6.1696E+02 9.2640E+02; 5.0800E+00 3.7938E+02 4.9389E+02 7.4175E+02; 5.0866E+00 3.1395E+02 4.0881E+02 6.1411E+02;...
    5.0952E+00 2.5389E+02 3.3072E+02 4.9695E+02; 5.1084E+00 2.0633E+02 2.6891E+02 4.0425E+02; 5.1210E+00 1.8641E+02 2.4309E+02 3.6563E+02;...
    5.1351E+00 1.8826E+02 2.4568E+02 3.6974E+02; 5.1482E+00 2.2679E+02 2.9615E+02 4.4595E+02; 5.1550E+00 2.6043E+02 3.4023E+02 5.1252E+02;...
    5.1595E+00 2.9156E+02 3.8102E+02 5.7412E+02; 5.1642E+00 3.2034E+02 4.1875E+02 6.3115E+02; 5.1682E+00 3.4519E+02 4.5139E+02 6.8055E+02;...
    5.1781E+00 3.5957E+02 4.7045E+02 7.0961E+02; 5.1847E+00 3.4603E+02 4.5298E+02 6.8357E+02; 5.1954E+00 3.2944E+02 4.3156E+02 6.5164E+02;...
    5.2055E+00 3.3366E+02 4.3739E+02 6.6085E+02; 5.2138E+00 3.4914E+02 4.5798E+02 6.9234E+02; 5.2208E+00 3.6640E+02 4.8091E+02 7.2738E+02;...
    5.2296E+00 3.9177E+02 5.1453E+02 7.7865E+02; 5.2354E+00 4.0550E+02 5.3286E+02 8.0679E+02; 5.2433E+00 4.0756E+02 5.3590E+02 8.1185E+02;...
    5.2517E+00 3.9166E+02 5.1534E+02 7.8116E+02; 5.2587E+00 3.5831E+02 4.7180E+02 7.1558E+02; 5.2683E+00 3.3155E+02 4.3692E+02 6.6315E+02;...
    5.2789E+00 3.2360E+02 4.2684E+02 6.4839E+02; 5.2903E+00 3.3091E+02 4.3692E+02 6.6427E+02; 5.3008E+00 3.4323E+02 4.5364E+02 6.9029E+02;...
    5.3092E+00 3.5956E+02 4.7569E+02 7.2447E+02; 5.3234E+00 3.8106E+02 5.0478E+02 7.6962E+02; 5.3377E+00 3.9959E+02 5.3005E+02 8.0911E+02;...
    5.3513E+00 3.9875E+02 5.2966E+02 8.0948E+02; 5.3630E+00 3.7824E+02 5.0308E+02 7.6973E+02; 5.3769E+00 3.5212E+02 4.6900E+02 7.1847E+02;...
    5.3905E+00 3.4121E+02 4.5513E+02 6.9810E+02; 5.4047E+00 3.5998E+02 4.8086E+02 7.3851E+02; 5.4164E+00 3.9804E+02 5.3242E+02 8.1865E+02;...
    5.4294E+00 4.4348E+02 5.9404E+02 9.1453E+02; 5.4420E+00 4.6787E+02 6.2764E+02 9.6748E+02; 5.4561E+00 5.0167E+02 6.7410E+02 1.0000E+03;...
    5.4776E+00 5.7959E+02 7.8040E+02 1.0000E+03; 5.4947E+00 6.1694E+02 8.3219E+02 1.0000E+03; 5.5052E+00 5.7136E+02 7.7172E+02 1.0000E+03;...
    5.5136E+00 4.9816E+02 6.7363E+02 1.0000E+03; 5.5271E+00 4.4811E+02 6.0678E+02 9.4299E+02; 5.5410E+00 4.6541E+02 6.3111E+02 9.8203E+02;...
    5.5519E+00 4.9859E+02 6.7697E+02 1.0000E+03; 5.5651E+00 5.1011E+02 6.9348E+02 1.0000E+03; 5.5729E+00 4.7243E+02 6.4285E+02 1.0000E+03;...
    5.5782E+00 4.2846E+02 5.8341E+02 9.1107E+02; 5.5839E+00 3.7967E+02 5.1732E+02 8.0833E+02; 5.5912E+00 3.3024E+02 4.5032E+02 7.0413E+02;...
    5.5993E+00 2.9970E+02 4.0905E+02 6.4012E+02; 5.6121E+00 2.8073E+02 3.8363E+02 6.0097E+02; 5.6242E+00 2.8804E+02 3.9412E+02 6.1807E+02;...
    5.6379E+00 3.1806E+02 4.3571E+02 6.8403E+02; 5.6452E+00 3.5199E+02 4.8261E+02 7.5821E+02; 5.6516E+00 3.9121E+02 5.3673E+02 8.4372E+02;...
    5.6562E+00 4.2596E+02 5.8470E+02 9.1953E+02; 5.6607E+00 4.6848E+02 6.4338E+02 1.0000E+03; 5.6671E+00 5.3968E+02 7.4157E+02 1.0000E+03;...
    5.6720E+00 6.0503E+02 8.3175E+02 1.0000E+03; 5.6763E+00 6.8668E+02 9.4437E+02 1.0000E+03; 5.6784E+00 7.1880E+02 9.8883E+02 1.0000E+03;...
    5.6814E+00 7.7750E+02 1.0000E+03 1.0000E+03; 5.6885E+00 9.0768E+02 1.0000E+03 1.0000E+03; 5.7000E+00 1.0000E+03 1.0000E+03 1.0000E+03;...
    5.7126E+00 1.0000E+03 1.0000E+03 1.0000E+03; 5.7502E+00 1.0000E+03 1.0000E+03 1.0000E+03; 5.7700E+00 1.0000E+03 1.0000E+03 1.0000E+03;...
    5.7818E+00 9.3692E+02 1.0000E+03 1.0000E+03; 5.7870E+00 8.5000E+02 1.0000E+03 1.0000E+03; 5.7911E+00 7.7051E+02 1.0000E+03 1.0000E+03;...
    5.7940E+00 7.1994E+02 1.0000E+03 1.0000E+03; 5.7961E+00 6.8831E+02 9.5629E+02 1.0000E+03; 5.8005E+00 6.0464E+02 8.4040E+02 1.0000E+03;...
    5.8136E+00 4.4000E+02 6.1209E+02 9.7390E+02; 5.8350E+00 2.8686E+02 3.9954E+02 6.3640E+02; 5.8621E+00 2.2484E+02 3.1360E+02 5.0013E+02;...
    5.8921E+00 2.0016E+02 2.7956E+02 4.4637E+02; 5.9327E+00 2.0486E+02 2.8647E+02 4.5789E+02; 5.9656E+00 2.2016E+02 3.0810E+02 4.9278E+02;...
    6.0280E+00 2.5887E+02 3.6238E+02 5.7977E+02; 6.0769E+00 3.1031E+02 4.3443E+02 6.9508E+02; 6.1395E+00 4.1120E+02 5.7568E+02 9.2108E+02;...
    6.1600E+00 4.5786E+02 6.4100E+02 1.0000E+03; 6.1823E+00 5.2017E+02 7.2824E+02 1.0000E+03; 6.2000E+00 5.8714E+02 8.2200E+02 1.0000E+03;...
    6.2125E+00 6.4071E+02 8.9700E+02 1.0000E+03; 6.2214E+00 6.8356E+02 9.5699E+02 1.0000E+03; 6.2267E+00 7.1071E+02 9.9500E+02 1.0000E+03;...
    6.2334E+00 7.5000E+02 1.0000E+03 1.0000E+03; 6.2498E+00 8.4286E+02 1.0000E+03 1.0000E+03; 6.2650E+00 9.5000E+02 1.0000E+03 1.0000E+03;...
    6.2800E+00 1.0000E+03 1.0000E+03 1.0000E+03; 6.3103E+00 1.0000E+03 1.0000E+03 1.0000E+03; 6.3500E+00 1.0000E+03 1.0000E+03 1.0000E+03;...
    ];
end
end //end of create global TableOfBBOabsorptionCoef_iwfjp

//absorption coefficient calculation
AbsCoef_min=interp1(TableOfBBOabsorptionCoef_iwfjp(:,1), TableOfBBOabsorptionCoef_iwfjp(:,2), inputValues, 'linear', 'extrap');
AbsCoef_typ=interp1(TableOfBBOabsorptionCoef_iwfjp(:,1), TableOfBBOabsorptionCoef_iwfjp(:,3), inputValues, 'linear', 'extrap');
AbsCoef_max=interp1(TableOfBBOabsorptionCoef_iwfjp(:,1), TableOfBBOabsorptionCoef_iwfjp(:,4), inputValues, 'linear', 'extrap');
// end of absorption coefficient calculation

//refraction index calculation
RefracInd_o=sqrt(1+ inputValues.^2 *0.90291 ./ (inputValues.^2 -0.003926) +...
 inputValues.^2 *0.83155 ./ (inputValues.^2 -0.018786) +...
 inputValues.^2 *0.76536 ./ (inputValues.^2 -60.01));

RefracInd_e=sqrt(1+ inputValues.^2 *1.151075 ./ (inputValues.^2 -0.007142) +...
 inputValues.^2 *0.21803 ./ (inputValues.^2 -0.02259) +...
 inputValues.^2 *0.656 ./ (inputValues.^2 -263));
// end of refraction index calculation

// data to return
ValueToReturn=[inputValues, AbsCoef_min, AbsCoef_typ, AbsCoef_max, RefracInd_o, RefracInd_e];

endfunction
```

An example for use:

```
SomeData=AbsorptionBBO_v0(0.30:0.002:2.0); //getting data for 300-2000 nm
scf(287);clf();
subplot(3,1,1);
 plot('nl',SomeData(:,1),SomeData(:,2:4)); xgrid;
 xtitle('', 'Wavelength, mkm', 'Attenuation, -dB/cm');
subplot(3,1,2);
InPower=2; //input power W (no reflection losses)
CrystalLength=5.5;//crystal length  mm
 plot('nl',SomeData(:,1), InPower.*(1-10.^(-SomeData(:,2:4)/10.* CrystalLength/10 )) );
 xtitle('Absorbed power: '+string(InPower)+'W input, length '+…
  string(CrystalLength)+'mm', 'Wavelength, mkm', 'Absorbed power, W'); xgrid;
subplot(3,1,3);
 plot(SomeData(:,1),SomeData(:,5:6));
 xtitle('', 'Wavelength, mkm', 'Refractive index');xgrid;
```

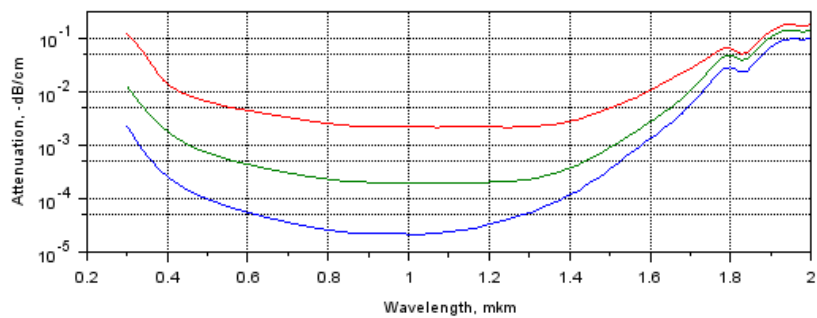

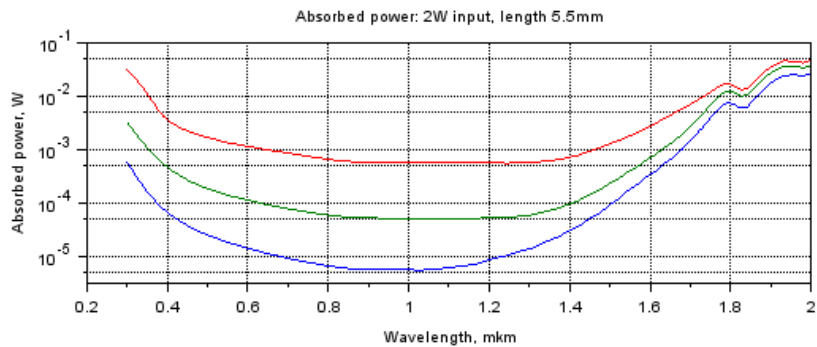

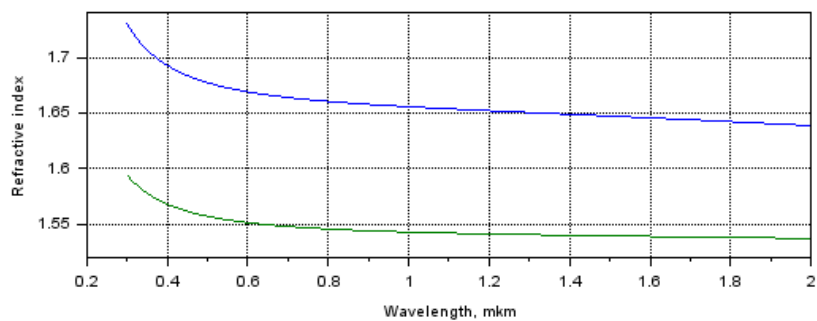